\documentclass[aps,pre,twocolumn,showpacs,groupedaddress,floatfix]{revtex4}

\usepackage[latin1]{inputenc} 
\usepackage{fancyhdr}         
\usepackage{appendix}

\usepackage{graphicx}
\usepackage{dcolumn}
\usepackage{bm}
\usepackage{color}

\newcommand{\mups}{$~\mu$m$\cdot$s$^{-1}$}

\begin{document}

\title{Boundary-induced inhomogeneity of particle layers in the solidification of suspensions}

\author{Brice Saint-Michel$^{1,3}$, Marc Georgelin$^1$,  Sylvain Deville$^2$, and Alain Pocheau$^1$}
\affiliation{
$^1$Aix-Marseille Univ, CNRS, Centrale Marseille, IRPHE, Marseille, France\\
$^2$Laboratoire de Synth\`ese et Fonctionnalisation des C\'eramiques, UMR3080 CNRS/Saint-Gobain CREE, Saint-Gobain Research Provence, Cavaillon, France\\
$^3$Now at: Department of Chemical Engineering, Imperial College London, London SW7 2AZ , United Kingdom
}
%


\date{\today}

\begin{abstract}
When a suspension freezes, a compacted particle layer builds up at the solidification front with noticeable implications on the freezing process. 
In a directional solidification experiment of monodisperse suspensions in thin samples, we evidence a link between the thickness of this layer and the sample depth.
We attribute it to an inhomogeneity of particle density that is attested by the evidence of crystallization at the plates and of random close packing far from them.
A mechanical model based on the resulting modifications of permeability enables us to relate the layer thickness to this inhomogeneity and to select the distribution of particle density that yields the best fit to our data.
This distribution involves an influence length of sample plates of about eleven particle diameters.
Altogether, these results clarify the implications of boundaries on suspension freezing.
They may be useful to model polydisperse suspensions with large particles playing the role of smooth boundaries with respect to small ones.
\end{abstract}


\maketitle

\section{Introduction}

The solidification of suspensions is a phenomenon that appears both in nature and in dedicated applications.
In nature, repeated freezing/thawing cycles induce frost heave \cite{Zhu2000,Rempel2004,Peppin2013}, ice lens formation \cite{Mutou1998,Saruya2013,Anderson2014} or cryoturbation \cite{Ping2008} whose implications on soils are responsible for costly damages to roads, buildings or manmade structures.
Applications address food engineering \cite{Velez-Ruiz2007}, cryobiology \cite{Bronstein1981,Korber1988,Muldrew2000} or the fabrication of materials to obtain particle-reinforced alloys by casting \cite{Stefanescu1988} or bio-inspired porous or composite materials by freezing \cite{Deville2007a}.
The quest for understanding the mechanisms at work in these different processes has stimulated a number of studies dedicated to the interaction of single \cite{Uhlmann1964,Cisse1971,Zubko1973,Chernov1976,Korber1985,Lipp1990,Lipp1993,Rempel1999,Rempel2001,Park2006} or multiple particles \cite{Dash2006,Peppin2006,Peppin2008,Anderson2012,Anderson2014,Saint-Michel2017} with a solidification front.

In particular, in a number of situations, the front velocity is too slow to trap an isolated particle.
A compacted particle layer then develops ahead of the front until trapping conditions are eventually reached \cite{Anderson2012,Saint-Michel2017} and make the layer stop growing.
The mechanical features, the organization and the interaction of this layer with the solidification front are essential to predict or uncover the global evolution of a freezing suspension.
However, suspensions are usually considered in an unlimited space whereas some degree of confinement may be present in practice due to system boundaries or to inclusions of additional elements of large size compared to particles (e.g. gravels or rocks).
Considering the influence of space confinement on suspension solidification may thus provide valuable information for material processing or for modeling the freezing of composite suspensions.
We address this issue here by using the availability of varying the suspension depth of thin samples in directional solidification.

Changing the depth of the samples in which the directional freezing of monodisperse suspensions is studied, we evidence, at any solidification velocity, a variation of the particle layer thickness with the sample depth.
On the other hand, observation of particles close to the sample plates reveals an hexagonal lattice configuration that differs from the random close packing evidenced far away.
This results in a variation of particle volume fraction along the sample depth whose implication on the particle layer thickness is determined using a mechanical model of trapping and repelling forces on particles adjacent to the solidification front.
Approximating the particle layers at the smallest and largest depths as homogeneous, respectively fully crystallized and random close packed, we show that their change of permeability explains their change of layer thickness.
For intermediate sample depths, we consider one or two parameters models of the evolution of particle volume fraction from the plates to the bulk.
This enables us to confront these models to our experimental data and to select the best fitting particle density evolution.
This yields us to recover the evolution of the layer thickness with the sample depth and to refine the determination of the mean thermomolecular pressure exerted by a solidification front on particles during their trapping. 
It should be \emph{a priori} possible to extend these determinations to any particle size and any suspension.

Section \ref{Experiment} describes the experiment setup and the generic evolution with the solidification velocity of the particle layer thickness. 
Section \ref{ParticleTrappingPT} first establishes the link between this evolution and the repelling thermomolecular pressure exerted by the solidification front on nearby particles.
It then reports the different evolutions measured for various sample depths.
Section \ref{ParticleLayerInhomogeneity} addresses the origin of the inhomogeneity of particle density in the particle layer and its mechanical implication on trapping particles.
Simple models of particle density are then considered to recover the experimental variations with sample depth.
A discussion and a conclusion about the study follow.

\section{Experiment}
\label{Experiment}

\subsection{Setup}
\label{Setup}

\begin{figure*}[htbp]
\includegraphics[width=8cm]{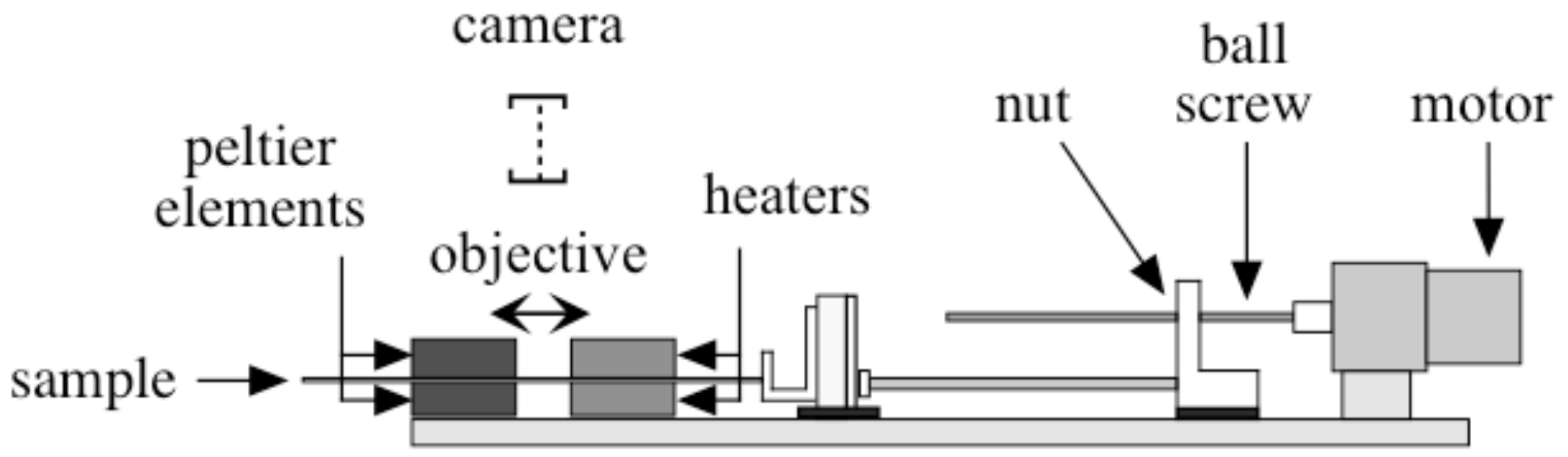}
\hspace{0.5cm}
\includegraphics[width=8cm]{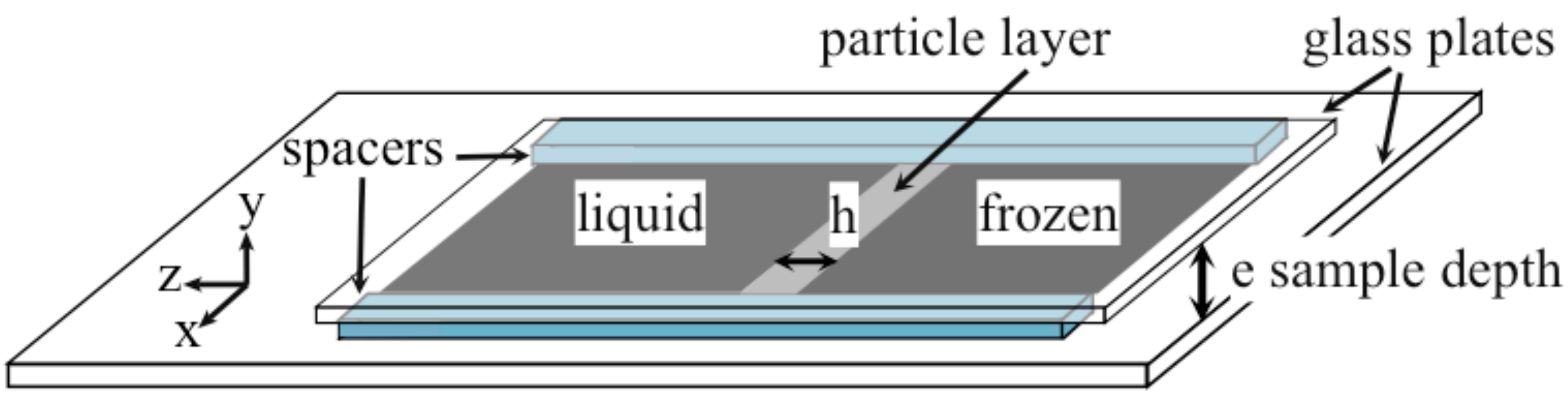}
\\
	\flushleft{\hspace{4cm} (a) \hspace{8cm} (b)}
\caption{(Color online)
Sketch of the experimental setup. 
(a) A micro-stepper motor coupled to a linear track pushes a thin sample in between heaters and coolers so as to force it to solidify under controlled conditions.
An optical access enables a real-time non-invasive visualization of the vicinity of the solidification interface.
It is composed of an objective which makes the image of the sample on a camera.
(b) Samples are made of two glass plates separated by calibrated spacers.
They provide a large domain, $100$ mm long and $45$ mm wide, filled with the suspension to solidify and whose thickness $e$ can be varied by the spacers. 
Plates are sealed so as to enclose the suspension.
When a suspension freezes, a compacted particle layer of thickness $h$ builds up ahead of the solidification front. The axes $x$, $y$ and $z$ refer to the directions of the solidification front, the sample depth and the thermal gradient respectively.
}
\label{Set-up}
\end{figure*}

The experimental setup aims at achieving the directional solidification of a thin sample under controlled conditions while allowing the visualization of the vicinity of the solidification interface.
It consists in pushing at a definite velocity a sample in a uniform thermal gradient [Fig.\ref{Set-up}(a)], following the Bridgman-Stockbarrer technique \cite{Bridgman1925,Stockbarger1936} and its application to thin samples \cite{Hunt1966}.
Here, the present setup was originally conceived for the directional solidification of binary mixtures \cite{Georgelin1998,Pocheau2006,Deschamps2006} and recently applied to the solidification of suspensions \cite{Saint-Michel2017}.
Instead of varying the thermal field \cite{Saruya2013,Korber1985,Cisse1971}, it thus varies the sample position in a fixed thermal field in samples thinner than in refs. \cite{Mutou1998,Peppin2008,Anderson2012,Anderson2014} and larger than in refs. \cite{Lipp1990,Lipp1993,Beckmann1990}.

The sample translation is obtained from a screw rotated at a controlled rate by a microstepper motor (ESCAP).
Thanks to a recirculating ball screw (Transroll), this rotation induces a regular translation of a sample holder on a linear track (THK).
With $6400$ microsteps by turn and a $5$mm screw pitch, the elementary displacement is $0.8 \mu$m.
Vibration at the end of micro-displacements are minimized by the use of an electronic damping to slow down the motor rotation.
Velocities up to $50~\mu$m.s$^{-1}$ can be achieved with relative modulations less than $3\%$.

A controlled thermal gradient is provided by heaters and coolers separated by a $10$ mm gap.
They are electronically regulated at temperatures of $\pm 20^{\circ}$C.
As these temperatures place the melting isotherm in the center of the gap, the visualization of the solidification interface is facilitated and the thermal gradient dependence on the sample velocity $V$ is minimized \cite{Georgelin1998,Pocheau2009}.
Both heaters and coolers involve copper blocks either heated by resistive sheets (Minco) or cooled by Peltier devices (Melcor).
To ensure a good thermal contact and the absence of inclined thermal gradient, the samples are sandwiched by top and bottom thermal blocks.
An external circulation of a cryogenic fluid at $- 30^{\circ}$C enables heat to be extracted from the Peltier devices and from the lateral sides of the setup.
The whole setup is finally surrounded by insulating polystyrene walls to provide a closed dry atmosphere that helps avoiding condensation and ice formation.

Samples are composed of two glass plates separated by 
calibrated propylene spacers [Fig.\ref{Set-up}(b)].
When held together, they delimit a parallelepipedic space in which the suspension is introduced by capillarity prior to sealing.
The plates dimensions, $100\times45\times0.7 \textrm{mm}^3$ for the top glass and $150\times50\times0.8 \textrm{mm}^3$ for the bottom glass, have been chosen large enough for providing a large central zone free of boundary disturbances.
The spacer thickness allows a variety of sample depth $e$.
Here, six depths were studied : $16, 30, 50, 75, 100$ and $125 \mu$m.

The suspensions contained plain polystyrene (PS) spheres of $3~\mu$m diameter and density $1.05$
at volume fraction $\phi_0=10\%$ or $20\%$. 
They were manufactured by Magsphere Inc. and were stable over months.
The standard deviation of their diameter, $0.12 \mu$m, yields a relative standard deviation of $4\%$. 
This results in a monomodal particle distribution with a low polydispersity, as confirmed by confocal microscopy [see Fig. \ref{Hexagonal}(b) in section \ref{PlateOrdering}].

Solutal effects were investigated by filtering out the particles using chromatography micro-filters and looking for the morphological instability of planar solidification fronts in the resulting mixture.
The large critical velocity then found, of several \mups{}, indicates a low concentration of additive.
The dynamical viscosity $\mu$ of the liquid contained in the suspension was thus taken as that of water : $\mu=1.8\times 10^{-3}$ Pa.s.

%

An optical access in the middle of the gap between heaters and coolers enables visualization of the vicinity of the solidification interface [Fig.\ref{Set-up}(a)].
In order to prevent the solidification front from perturbations, an exploded optical setup has been preferred to a microscope.
It is composed of a photographic lens of focal length $50$mm  placed at about this distance to the solidification front so as to provide an image of large magnification on a camera placed about a meter apart.
As the rays are weakly inclined, the Gauss approximation is fairly satisfied.
This guarantees stigmatism and thus an excellent image sharpness.

\begin{figure}[h]
\includegraphics[width=8.5cm]{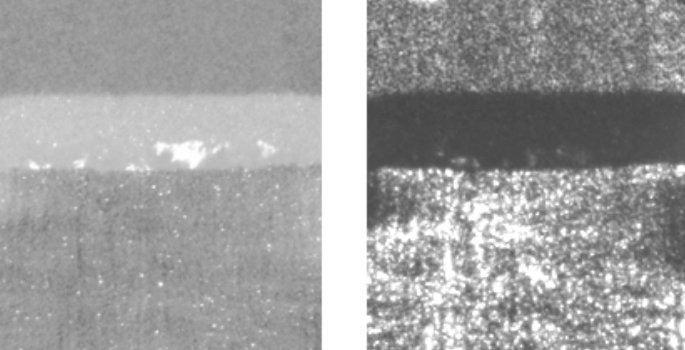}
\caption{
Comparison between the optical methods of reflection and transmission.
The same situation is observed by reflection (left) and by transmission (right).
It involves a top zone (the liquid phase), a bottom zone (the frozen phase) and an intermediate zone (the compacted particle layer).
The change of optical methods inverses bright and dark zones.
In particular, the top and bottom zones remain grey.
In contrast, the intermediate zone turns from bright to dark.
Interestingly, it keeps the same thickness whatever the optical method.
The bright patches displayed in the particle layer by reflection (left) correspond to the long-range particle ordering at the sample plates evidenced in figure \ref{Confocal}-b.
}
\label{ReflectionTransmission}
\end{figure}

As particles diffuse light, observation may be achieved either by reflection or transmission (Fig. \ref{ReflectionTransmission}).
In both cases, the intensity received depends on the particle volume fraction $\phi$ : low (resp. large) at large $\phi$ in transmission (resp. reflection).
Whereas both methods provided grey images on both the solid and liquid phases due to their moderate particle volume fraction ($\phi_0=10\%$ or $20\%$), the particle layer that forms in between at a much larger volume fraction ($\phi\approx0.64$) appeared either dark (transmission) or bright (reflection).
Interestingly, its apparent thickness remains the same whatever the optical method (Fig. \ref{ReflectionTransmission}).
Both of them could thus be used to document the particle layer thickness in the vicinity of plates by reflection or through the entire sample depth by transmission.
The reflection method has been the most applied in this study.
In complement, confocal microscopy (Leica SP8, combined to a Leica DM6000 optical microscope), used with a long working distance non-immersive objective (Leica HC PL APO 20x/0.70 CS) has also been used to determine the particle arrangement in the vicinity of the sample plates.

The directions of the solidification front, the sample depth and the thermal gradient will be taken as the $x$-axis, the $y$-axis and the $z$-axis respectively [Fig.\ref{Set-up}(b)].
The particle layer thus develops in the direction $z$, normal to the direction $y$ of the sample depth and extends along the $x$ direction up to the sample lateral limits.

\subsection{Particle layer thickness}
\label{Particle layer thickness}

\begin{figure}[ht!]
\centering
\includegraphics[width=8.5cm]{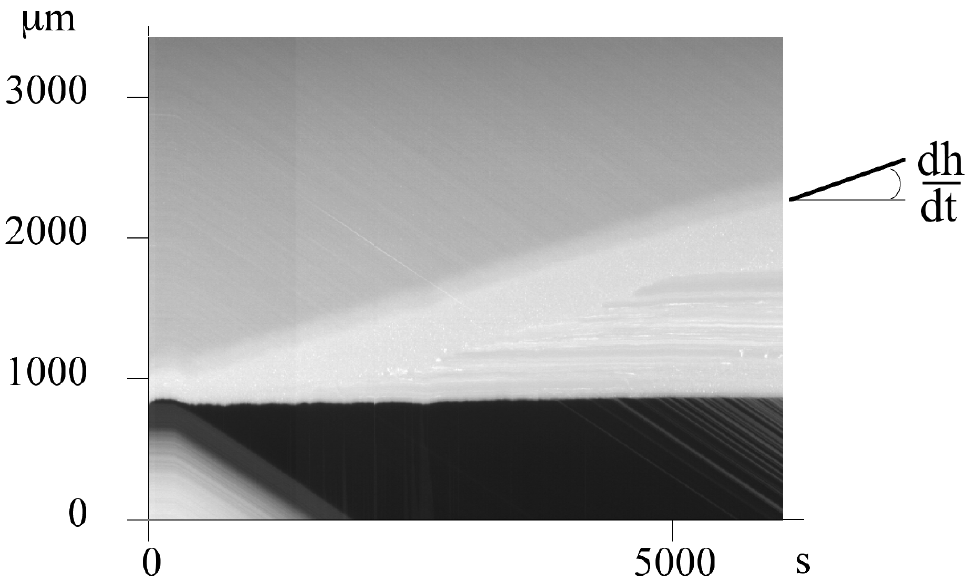}
\caption{Spatio-temporal diagram of the growth of the particle layer : $\phi_0 = 20\%$, $V=0.5$\mups{}, $e=125 \mu$m.
Visualization is performed by reflection.
The horizontal axis displays time.
The vertical axis displays a cross-section of the suspension from the frozen phase to the liquid phase at a fixed front position and at successive times.
The dark zone corresponds to the solid phase, the bright growing zone to the particle layer and the grey zone above it to the suspension.
The solidification front thus stands at the transition between the dark zone and the bright zone.
The darkness of the solid phase shows that no particle enter it during the building-up of the layer.
The growth rate $dh/dt$ of the layer thickness then yields, by particle conservation, a mean particle volume fraction $\phi_l=0.634$ that is equal to that of close random packing.
}
\label{DST}
\end{figure}

When a sample starts solidifying, particles are first repelled by the solidification front.
They then accumulate ahead of it in a particle layer which involves a large particle volume fraction $\phi$ (Fig. \ref{ReflectionTransmission}).
By particle conservation, the growth rate of its thickness $h(t)$ provides the opportunity to determine its mean particle volume fraction $\phi_l$.
As the frontier F between the suspension and the particle layer advances at velocity $\mathbf{V}_F= dh/dt \; \mathbf{e}_z$ in the front frame, particles arrive on it at velocity $\mathbf{V}_P=  -V \mathbf{e}_z - \mathbf{V}_F$, i.e., $\mathbf{V}_P=  - (V+dh/dt) \mathbf{e}_z$.
As no particle enters the solid phase, the particle balance in the layer then yields $\phi_l dh/dt = \phi_0 (V+dh/dt)$ and finally $\phi_l = \phi_0 [ 1+ V/(dh/dt) ]$.
Figure \ref{DST} shows a spatio-temporal diagram of the building-up of the particle layer at  the largest sample depth, $e=125 \mu$m.
The growth rate of the layer thickness then provides $\phi_l=0.634\pm0.007$ \cite{Saint-Michel2017}, which is the value displayed by random close packing density in three dimensions 
$\phi_{rcp}=0.634$ \cite{Song2008}.
Hereafter, we shall denote this density $\phi_3$ to emphasize that it refers to a three-dimensional (3D) space.
The equality $\phi_l \approx \phi_3$ then means that the built-up layer is both random and compacted.

\begin{figure}[t!]
\centering
\includegraphics[width=8.5cm]{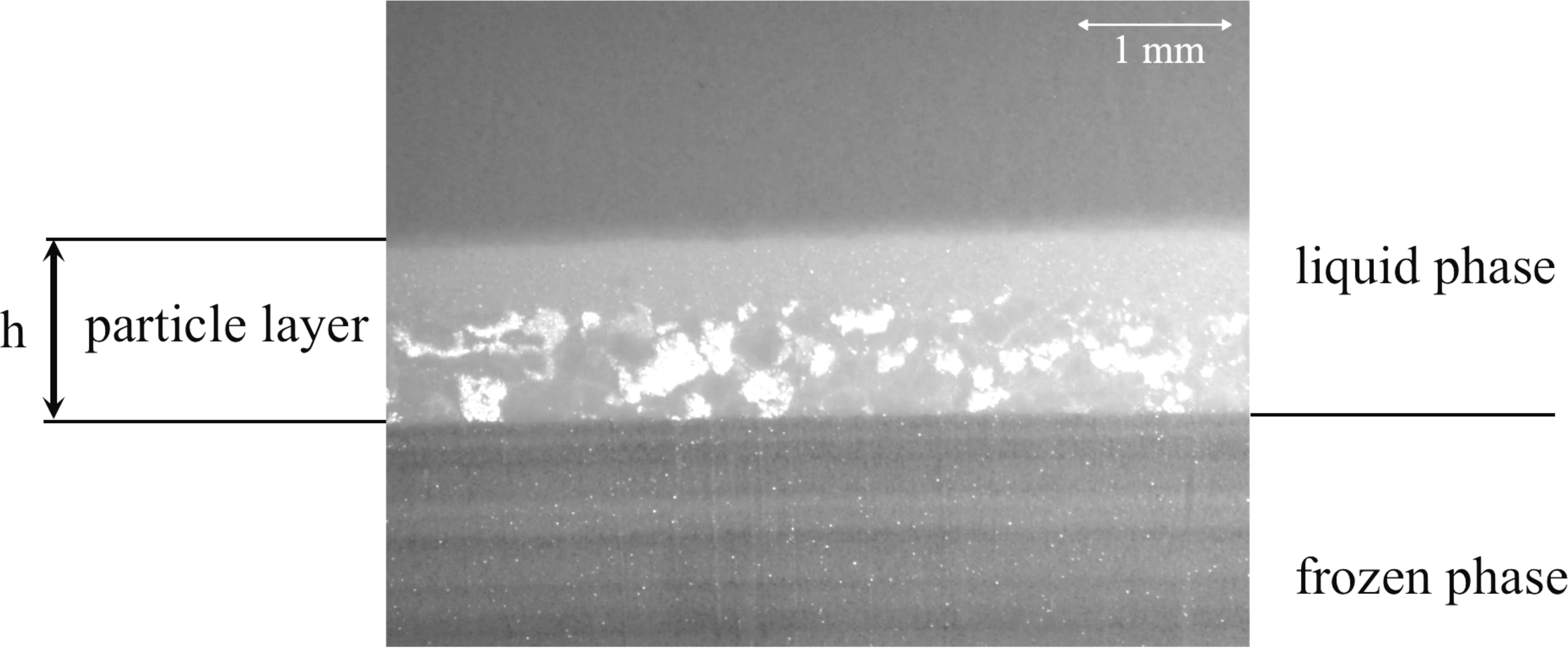}
\caption{
Reflection image of a compacted layer of particles formed ahead of the solidification front : $\phi_0 = 20\%$ , $e=50 \mu$m, $V=2$\mups{}.
The top and bottom of the image respectively correspond to the liquid phase and the frozen phase. 
They appear grey as the particle volume fraction, $\phi_0$, is low.
In between them, the bright zone reveals a noticeable increase of particle volume fraction in the compacted particle layer.
Its bright patches correspond to the long-range particle ordering evidenced at the sample plates in figure \ref{Confocal}-b.
The steady state thickness reached by the particle layer beyond its build-up is labeled $h$. 
}
\label{CompactLayer}
\end{figure}

The occurrence of a compacted layer largely increases the hydrodynamic viscous dissipation of the suspension. 
As discussed in section \ref{ForceBalanceModel}, this results in large stresses pushing the particles adjacent to the front towards it, thus promoting particle inclusion in the solid matrix.
At some value $h$ of the particle layer thickness, particles are then no longer repelled but trapped by the solidification front.
The particle layer thus ceases to grow so that $h$ stands as its steady state thickness after the initial growth transient \cite{Saint-Michel2017} (Fig. \ref{CompactLayer}).
Its value, of the order of a millimeter, displays the specificity of being both mesoscopic and related to the microscopic trapping mechanism of particles by the front.
This makes $h$ a variable both easy to measure accurately and valuable for investigating the trapping mechanism.
We shall thus dedicate the remainder of the study to it.

The layer thickness $h$ \emph{a priori} depends on all the parameters of the study, especially the particle diameter $d$, the volume fraction $\phi_0$ of the suspension, that $\phi_l$ of the compacted layer, the solidification velocity $V$ and the sample depth $e$.
In particular, for a given suspension, it is found to decrease with the solidification velocity $V$, as displayed in figure \ref{G,h,V,U}(a) for $d=3 \mu$m, $e=125 \mu$m and either $\phi_0=0.1$ or $0.2$.

\begin{figure}[t!]
\includegraphics[width=6cm]{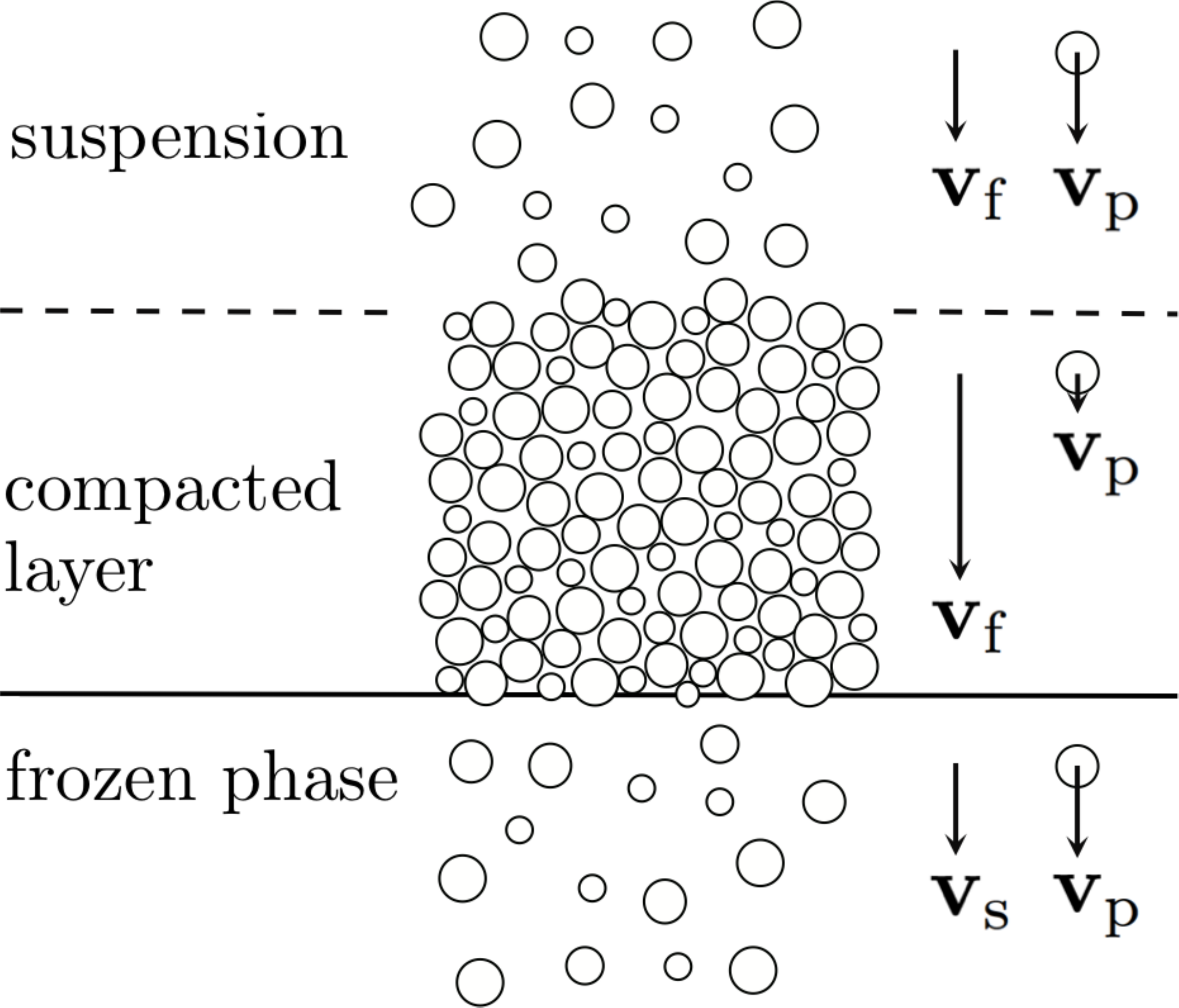}
\caption{Cross-section of the system showing the suspension, the compacted layer of particles and the frozen phase.
The mean velocities of the fluid, of the particles and of the solid with respect to the solidification front are denoted $\mathbf{v}_{\rm f}$, $\mathbf{v}_{\rm s}$ and $\mathbf{v}_{\rm p}$ respectively.
In the suspension and the frozen phase, they both equal the opposite of the solidification velocity. 
However, in the compacted layer of particles, they differ following the rise of particle volume fraction.
As particles are randomly distributed, their section by a plane displays here different radii although they actually have the same radius.
}
\label{Suspension}
\end{figure}

Although the two particle volume fractions $\phi_0$ display a similar trend, their data curves show noticeable differences [Fig. \ref{G,h,V,U}(a)].
This may be attributed to a wrong choice of velocity.
Indeed, as the layer thickness is related to viscous dissipation, the relevant flow velocity refers to the volume flux of liquid through the particle matrix.
This velocity $\mathbf{U}$, called the Darcy velocity, may be determined by considering the mean velocities of fluid $\mathbf{v}_{\rm f}$ and of particles $\mathbf{v}_{\rm p}$ in the front frame, all of them being directed along the $z$-axis.
Mass conservation in the compacted layer yields, for constant $h$, $v_{\rm f} = -  V (1-\phi_0)/(1-\phi_l)$ and $v_{\rm p} = -  V \phi_0/\phi_l$ (Fig.~\ref{Suspension}).
Whereas these velocities are equal in the incoming suspension, they thus differ in the particle layer, which generates viscous dissipation.
In particular, their difference corresponds to a volume flux of liquid $U$ with respect to the particle matrix equal to $U = (1-\phi_l) (v_{\rm f} - v_{\rm p})$ or :
\begin{equation}
\label{U}
U = - V \frac{(\phi_l-\phi_0)}{\phi_l}.
\end{equation} 
This Darcy velocity should therefore be more relevant to refer to the particle layer thickness $h$.
This is apparent in figure \ref{G,h,V,U}(b) where the same data no longer display differences regarding the particle volume fraction $\phi_0$ when $|U|$ is used instead of $V$.

As $\phi_l>\phi_0$, $U$ is negative so that a flow feeds the solidification front and allows the solid phase to advance.
For this reason, the intensity of the Darcy velocity will be hereafter denoted $|U|$.

The graph $h(1/|U|)$ displays a linear part up to $1/|U| \lesssim 1 s.\mu$m$^{-1}$ followed by a much slower increase yielding a noticeable concavity [Fig. \ref{G,h,V,U}(b)].
The linear part corresponds to $h \propto |U|^{-1}$.
This exponent differs from the value $-0.72$ reported by Anderson and Worster for alumina suspensions in similar conditions \cite{Anderson2012}.
This difference may be due to the polydispersity of the alumina suspension used in their study or to their closeness to the transition to an ice lens regime where $h$ is no longer steady.

The above change of trend indicates a change of the type of dominant dissipation in the system made by the particles, the fluid and the glass plates.
As the particle layer is uniformly pushed by the front, it involves no internal shear. Therefore, two kinds of dissipation may be invoked : (i) the viscous dissipation exerted by the fluid on the particles and the plates ; (ii) the solid friction exerted between the particles and the sample plates.

Viscous dissipation yields a pressure at the solidification front \emph{proportional} to the layer thickness $h$.
In contrast, solid friction between particles and boundaries is known in granular materials to induce, by the Janssen effect \cite{Janssen1895,JanssenSperl1895,AndreottiForterrePouliquen2013}, the pressure at  the bottom of granular columns to exponentially relax to a definite value as the column height $h$ increases.
This makes their apparent weight bounded to the actual weight of a length $\lambda$ of the columns, $\lambda$ being the Janssen's length.

Here, in the present context of suspension freezing, we have shown in a companion paper \cite{Saint-Michel2018} that viscous friction plays the role of gravity and that solid friction induces a similar {\it exponential} amplification of the particle pressure at the solidification front.
Both result in a universal relationship between $h$ and $U$ that remains the same independently of the nature of the dominant dissipation mechanism.
It can then be used to study the relation between $h$ and $U$ in our whole data set.

In the next section, after recalling the physical basis of this universal relationship, we use it to determine, at various sample depths $e$, the mean thermomolecular pressures $\bar{P}_T(e)$ exerted by solidification fronts on particles that enter it.
Their variations with $e$ will then lead us to question the role of the inhomogeneity of particle volume fraction in the layer behavior.

\begin{figure*}[t!!!]
\centering
\includegraphics[height=5.5cm]{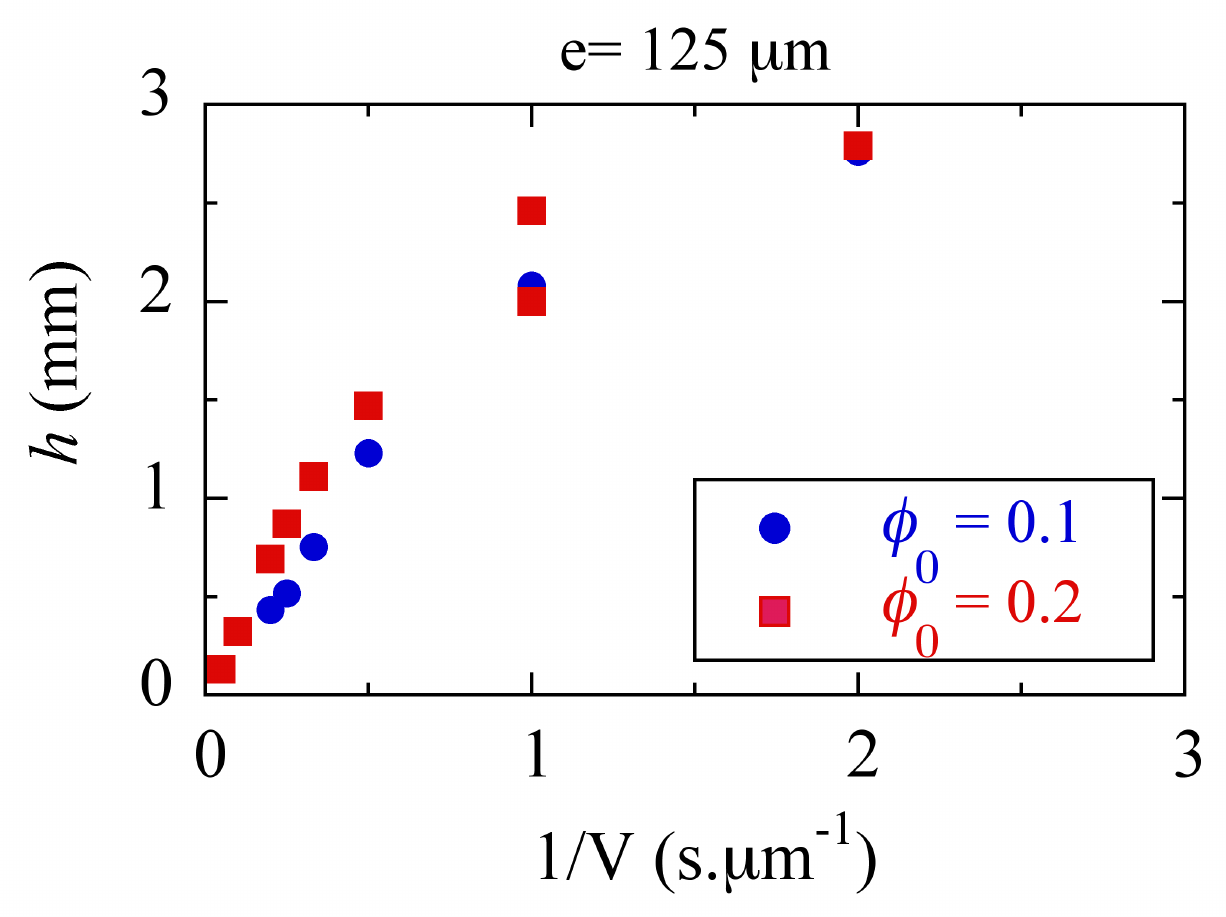}
\hspace{1cm}
\includegraphics[height=5.5cm]{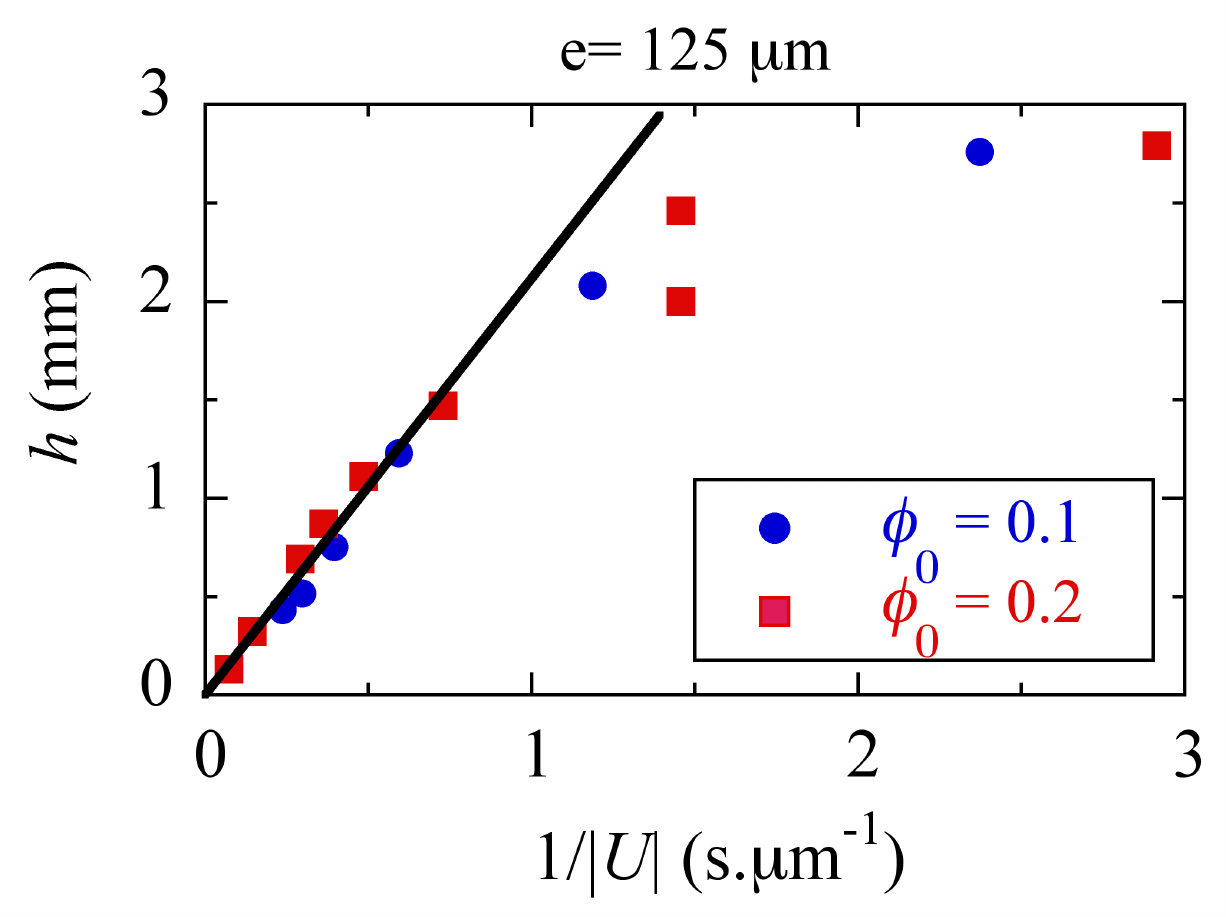}
\\
	\flushleft{\hspace{5cm} (a) \hspace{8cm} (b)}
\caption{(Color online)
Evolution of the steady state thickness $h$ of the compacted layer with the solidification velocity. 
The sample depth is $e=125 \mu$m and the particle volume fractions are $\phi_0=0.1$ and $0.2$. 
(a) Plot of the thickness $h$ with respect to the inverse of the velocity $V$. 
(b) Plot of the thickness $h$ with respect to the inverse of the Darcy velocity $U= -V (\phi_l-\phi_0)\phi_l$ with $\phi_l=\phi_3$.
The full line is a linear fit of data for $|U| > 1 \mu$m.s$^{-1}$.
}
\label{G,h,V,U}
\end{figure*}

\section{Particle trapping and thermomolecular pressure}
\label{ParticleTrappingPT}
The trapping of a particle by a solidification front is a phenomenon which obviously involves the interaction between both of them, at a small scale.
However, we shall see that the remaining particles of the layer, and hence, the whole particle matrix, also participate to it.
This offers the opportunity to indirectly measure, from the particle layer thickness $h$, the repelling thermomolecular force between particles and front.

\subsection{Force balance model}
\label{ForceBalanceModel}

\begin{figure}[t!!!]
\includegraphics[width=8cm]{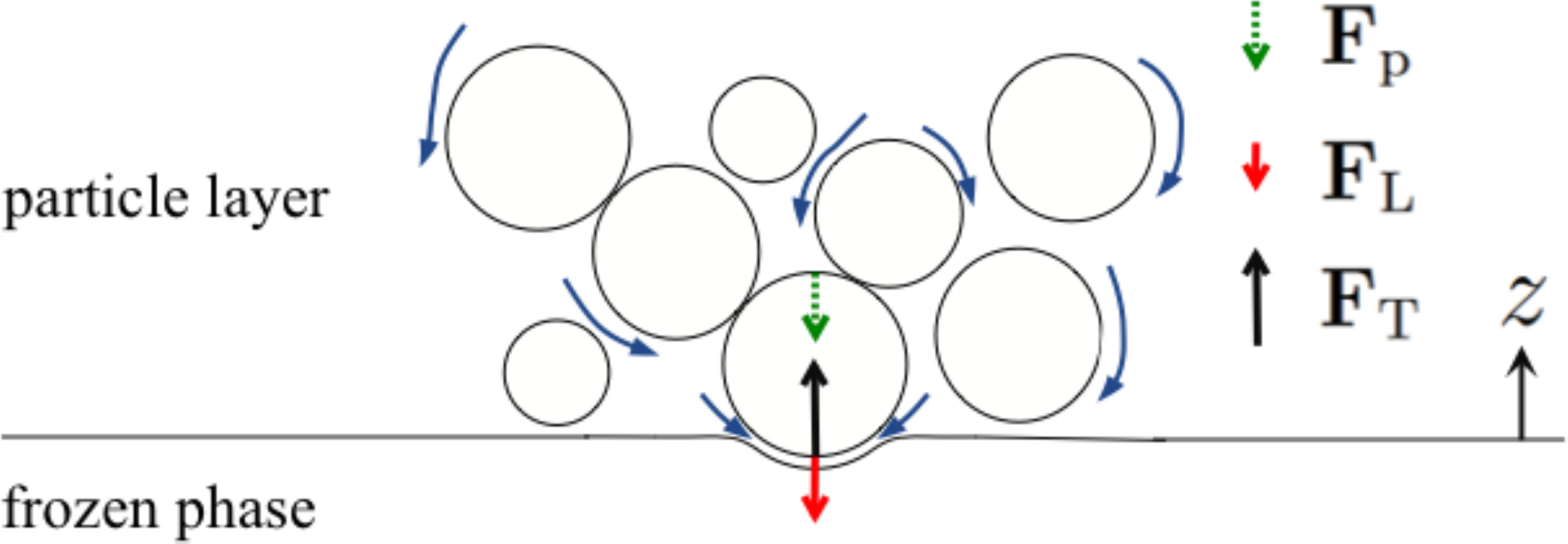}
\caption{(Color online)
Sketch of the forces acting on a particle entering the frozen phase: a thermomolecular repelling force $\mathbf{F}_{\rm T}$ exerted by the solidification front (black arrow); a lubrication force $\mathbf{F}_{\rm L}$ induced by the liquid flowing in the thin film which separates the particle and the front (red arrow); a force $\mathbf{F}_{\rm p}$ exerted by the particle layer on the particle, as a result of the viscous friction and the pressure drop induced by the liquid flow through the particle matrix (green dashed arrow). 
Blue curved arrows symbolize the liquid flow. 
The direction $\mathbf{e}_z$, normal to the solidification front, is parallel to the thermal gradient.
As particles are randomly distributed, their section by a plane displays different radii and few contact points here, although they actually have the same radius and are in contact with each other.
}
\label{ForcesParticuleEntrante}
\end{figure}

Three kinds of forces apply on a particle nearing a solidification front  (Fig. \ref{ForcesParticuleEntrante}) :

i) The thermomolecular force $\mathbf{F}_{\rm T}$ exerted by the solidification front on an entering particle. 

It results from van der Waals and electrostatic interactions between the particle and the front and stands here as a repelling force.
On an elementary particle surface, it corresponds to a normal force whose intensity quickly decreases with its distance to the front, as its inverse cube for non-retarded van der Waals interactions \cite{Israelachvili1992Book,Wilen1995} and as an exponential for electrostatic interactions \cite{Israelachvili1992Book,Wilen1995,Wettlaufer1999}.

On a spherical particle, the resultant of this force is, by symmetry, normal to the particle base, or equivalently parallel to the thermal gradient direction $\mathbf{e}_z$ (Fig.\ref{ForcesParticuleEntrante}). 
Its intensity depends on the distance between the particle base and the front.
However, as all particle layers stand in a state that corresponds to the trapping transition, we shall assume that this distance and thus the force intensity $F_T$ on a particle is a constant.
This will be corroborated below by the linear variation of the layer thickness $h$ with the inverse velocity $1/U$.

An important implication of the thermomolecular force is to induce an additional pressure between particles  and front which maintains a liquid phase between them, whatever the smallness of their distance \cite{Dash2006,Wettlaufer2006}.
Flows can thus occur in  these so-called premelted films, yielding a  lubrication force.

ii) The lubrication force $\mathbf{F}_{\rm L}$ on an entering particle.

 It results from viscous effects in the submicronic premelted film that separates an entering particle from the solidification front (Fig.\ref{ForcesParticuleEntrante}).
 Its resultant $\mathbf{F}_{\rm L}$ on a particle is by reason of symmetry parallel to the thermal gradient direction $\mathbf{e}_z$.
 As the corresponding flows around the particle are creeping flows, the intensity of $\mathbf{F}_{\rm L}$ is linearly related to their magnitude and thus to the Darcy velocity $U$ : $F_{\rm L} = f_L U$, the prefactor $f_L$ depending on the geometry of the film that separates a particle from the front (see refs. \cite{Rempel1999,Rempel2001,Park2006} for details).
 
iii) The force $\mathbf{F}_{\rm p}$  exerted by the particle layer on an entering particle.

 It results from the pressure drop and the viscous friction induced by the fluid flowing across the compacted particle layer and from the solid friction exerted by the plates on the sliding particles.
 It is transmitted to the particles nearing the front by contacts along the particle matrix.
 
We label $\underline{\sigma}^p$ the particle stress tensor.
Friction at the plates yields at $y=\pm e/2$, from the Coulomb's law, $\sigma^p_{yz}=\mu_W \sigma^p_{yy}$ between the normal stress $\sigma^p_{yy}$ and the tangential stress $\sigma^p_{yz}$, $\mu_W$ designing the friction coefficient between particles and plates.
In addition the redistribution of stresses inherent to granular materials \cite{Janssen1895,JanssenSperl1895} yields $\sigma^p_{yy}=K\sigma^p_{zz}$, $K$ designing the Janssen's redirection coefficient.
This provides altogether the stress boundary conditions : $\sigma^p_{yz}=\mu_W K \sigma^p_{zz}$ at $y=\pm e/2$.
   
As the suspension moves at a steady velocity, force equilibrium in it yields $\nabla \underline{\sigma}^p = \nabla p$ where $p$ designs the fluid pressure.
We denote by a tilde the averages on the $y$ direction.
Following the Hele-Shaw geometry of the sample, the sample width is large enough compared to the sample depth for allowing the stress $ \tilde{\sigma}^p_{zz}$ to be uniform in the $x$-direction, so that $\tilde{\sigma}^p_{zz} \equiv  \tilde{\sigma}^p_{zz}(z)$ here.
The stress relation completed by the frictional boundary conditions then yields, under usual assumptions invoked in granular materials \cite{AndreottiForterrePouliquen2013,deGennes1999}, the mean stress equation \cite{Saint-Michel2018} :
   \begin{equation}
\label{StressJanssen}
\partial_z \tilde{\sigma}^p_{zz} + 2 \frac{\mu_W K}{e} \tilde{\sigma}^p_{zz} = \frac{d\tilde{p}}{dz}.
\end{equation}
 
 The pressure gradient in the particle layer follows the Darcy law :
\begin{equation}
\label{Darcy}
\nabla p = - \mu \frac{\mathbf{U}}{k} = \mu \frac{|U|}{k} \mathbf{e}_z,
\end{equation}
where $\mu$ denotes the liquid viscosity, $k$ the medium permeability and $U$ the Darcy velocity.
 It induces a pressure drop between the solution and the front that is responsible for a flow of liquid towards the front, i.e. for cryosuction \cite{Wettlaufer2006}. 
Pressure and viscous forces then make the particle matrix push the particles close to the front towards it and promote their trapping.
On the other hand, according to the Kozeny-Carman relation, the permeability $k(\phi_l,d)$ of the particle layer depends both on its particle volume fraction $\phi_l$ and on the particle diameter $d$ \cite{AndreottiForterrePouliquen2013} :
  \begin{equation}
\label{KC}
k(\phi_l,d)=\frac{d^2}{180}  \frac{(1-\phi_l)^3}{\phi_l^2}.
\end{equation}
 
Integration of relation (\ref{StressJanssen}) from the front position $z=0$ to the end $z=h$ of the particle layer where $\tilde{\sigma}^p_{zz}(h)=0$ yields  :
\begin{equation}
\label{StressJanssenFront}
\tilde{\sigma}^p_{zz}(0) = - \frac{d\tilde{p}}{dz} \; \lambda \;[ \exp(\frac{h}{\lambda}) -1]
\end{equation}
with :
\begin{equation}
\label{lambda}
\lambda = \frac{e}{2 \mu_W K}.
\end{equation}

This exponential trend is akin to the Janssen effect in granular materials, the role of the leading force, gravity, being played here by the pressure gradient \cite{Janssen1895,JanssenSperl1895,AndreottiForterrePouliquen2013}.
However, friction forces are mobilized here in the same direction as the leading pressure gradient force instead of the opposite for granular columns in silos.
This results in an exponential enhancement of stress instead of an exponential relaxation in the latter case.
  
This stress applies on a surface of the suspension normal to the front. 
 However, as particles occupy only a proportion $\phi_l$ of it, the mean pressure $\bar{P_{\rm p}}$ on particles adjacent to the front reads as \cite{Saint-Michel2017,Saint-Michel2018} :
 \begin{equation}
\label{NormalizationPressure}
\bar{P_{\rm p}}=\frac{\tilde{\sigma}^p_{zz}(0)}{\phi_l}.
\end{equation}

Using the stress determination (\ref{StressJanssenFront}) and the Darcy law (\ref{Darcy}), we obtain \cite{Saint-Michel2018} :
\begin{equation}
\label{Pp,U}
\bar{P_{\rm p}}=- \frac{\mu|U|}{\phi_l k(\phi_l,d)} \; \lambda \; [ \exp(\frac{h}{\lambda}) -1].
\end{equation}

As the particle layer stands in a critical state for trapping, its thickness and the resulting pressure $\bar{P_{\rm p}}$ have grown up so as to just reach a force balance on particles nearing the front.
For convenience, we express this balance in terms of mean pressure rather than of forces, i.e. in terms of $z$-component of forces divided by the particle section $\pi d^2/4$.
For particles adjacent to the front, the above analysis then yields : $\bar{P_{\rm p}}+ \bar{P_{\rm L}} +\bar{P_{\rm T}} = 0$ where $\bar{P_{\rm p}}$, $\bar{P_{\rm L}}$ and $\bar{P_{\rm T}}$ denote the mean pressures exerted on an entering particle by the particle matrix, the premelted film (by lubrication) and the solidification front (by thermomolecular interactions).
Among these pressures, two of them $\bar{P_{\rm p}}$ and $\bar{P_{\rm L}}$ tend to induce trapping and are therefore negative. 
In contrast, the remaining thermomolecular pressure $\bar{P_{\rm T}}$ tends to repel particles and is thus positive.

However, at a velocity $U_c$, the  lubrication force is sufficient to induce particle trapping on a single particle.
No particle layer can then build up since all particles coming on the front are trapped without delay.
Accordingly $\bar{P_{\rm p}}=0$ and the force balance becomes :
$\bar{P_{\rm L}} +\bar{P_{\rm T}} = 0$ with $\bar{P_{\rm L}} = - g U_c$ and $g= + 4f_L /\pi d^2$.
This provides a link between the prefactor $g$ and $\bar{P_{\rm T}}$ which yields $\bar{P_{\rm L}} = -  \bar{P_{\rm T}} \; |U|/U_c$ and :
\begin{equation}
\label{ForcebalancePpPT}
\bar{P_{\rm p}} = - \bar{P_{\rm T}} \, (1 - \frac{|U|}{U_c}).
\end{equation}

Using the determination (\ref{Pp,U}) of $\bar{P_{\rm p}}$, we obtain the following relationship between the particle layer thickness $h$ and the thermomolecular pressure $\bar{P_{\rm T}}$ :
\begin{equation}
\label{h,1/U,1/Uc}
\frac{ \mu}{\phi_l k(\phi_l,d)} \; \lambda \; [\exp(\frac{h}{\lambda}) - 1] = \bar{P_{\rm T}} \, (\frac{1}{|U|} - \frac{1}{U_c}).
\end{equation}

The occurrence in this relation of $U$ (which depends on $\phi_0$) instead of $V$ (which is independent of it) explains the convergence of data on a master curve in figure \ref{G,h,V,U}(b) in contrast to figure \ref{G,h,V,U}(a).
In addition, relation (\ref{h,1/U,1/Uc}) provides an interesting connection between a macroscopic variable, $h$, and a microscopic one, the thermomolecular pressure $\bar{P_{\rm T}}$, that we shall exploit to evaluate the latter.

\subsection{Thermomolecular pressure}
\label{Thermomolecular pressure}

In reference \cite{Saint-Michel2018}, we already studied the evolution of the layer thickness $h$ in the same experiment and with the same data as those considered here.
We first noticed that a relevant common value of the critical velocity $U_c$ was $U_c=15 \mu$m.s$^{-1}$, close to the value $U_c=12 \mu$m.s$^{-1}$ \cite{Saint-Michel2017} expected from trapping models \cite{Rempel1999,Rempel2001,Park2006}.
We then demonstrated the relevance of relation (\ref{h,1/U,1/Uc}) with $\phi_l=\phi_3$ and free parameters $\bar{P_{\rm T}}$, $\lambda$.
In particular, we found that the best fitting values $\lambda(e)$ at  each sample depth $e$ were in average proportional to $e$, as expected from (\ref{lambda}) : $\lambda=10.0 \; e$.

This study also revealed a variation of the thermomolecular pressure $\bar{P_{\rm T}}$ with $e$ that we wish to clarify here.
For this, we reconsider relation (\ref{h,1/U,1/Uc}) with respect to the same data, still with $\phi_l=\phi_3$ following Sect. \ref{Particle layer thickness}, but with $\lambda$ now fixed to $\lambda=10.0 \; e$.

Figure \ref{G,PT,U,e} reports, for all the sample depths studied, the evolution with the inverse velocity $1/|U|$ of the left hand side of relation (\ref{h,1/U,1/Uc}) with $\phi_l=\phi_3$.
In agreement with relation (\ref{h,1/U,1/Uc}), all graphs show a linear trend.
Their corresponding fitted slope then provides, at each $e$, an indirect measure of $\bar{P_{\rm T}}$ on entering particles.

For layer thicknesses $h$ well below $\lambda$ , solid friction is negligible and relation (\ref{h,1/U,1/Uc}) reduces to a linear relationship between $h$ and $1/|U|$.
For $h$ larger than $\lambda$, solid friction is noticeable.
It then induces the concavity of the exponential $\exp(h/\lambda)$ which renders the relationship between $h$ and $1/|U|$ non-linear.

The ordinates at the transition value $h=\lambda$ are shown by dotted lines in figure \ref{G,PT,U,e}.
They delimit below a regime dominated by viscous dissipation
and above a regime dominated by solid friction.
This physical distinction however yields no implication for our purpose since the universal relationship (\ref{h,1/U,1/Uc}) holds for both regimes.

All the  linear fits obtained at various $e$ are synthesized in figure \ref{G,PT,U,all}.
As the thermomolecular pressure $\bar{P_{\rm T}}$ is set at the microscopic level across the thin premelted film that separates an entering particle from the front, one would expect it to be uncorrelated with macroscopic variables such as the sample depth $e$. 
However, figure \ref{G,PT,U,all} shows that the slopes of above linear relationships, and hence the resulting determinations of $\bar{P_{\rm T}}$, increase with the sample depth $e$.

This apparent dependence of the thermomolecular pressure $\bar{P_{\rm T}}$ on the sample depth $e$, $\bar{P_{\rm T}}\equiv \bar{P_{\rm T}}(e)$, therefore seems somewhat paradoxical.
However, we shall show in the next section that the sample depth $e$ actually enters the picture, not on the magnitude of the thermomolecular pressure $\bar{P_{\rm T}}$, but on the profile of the particle volume fraction in the layer.

\begin{figure*}[t!]
\centering
\includegraphics[height=6cm]{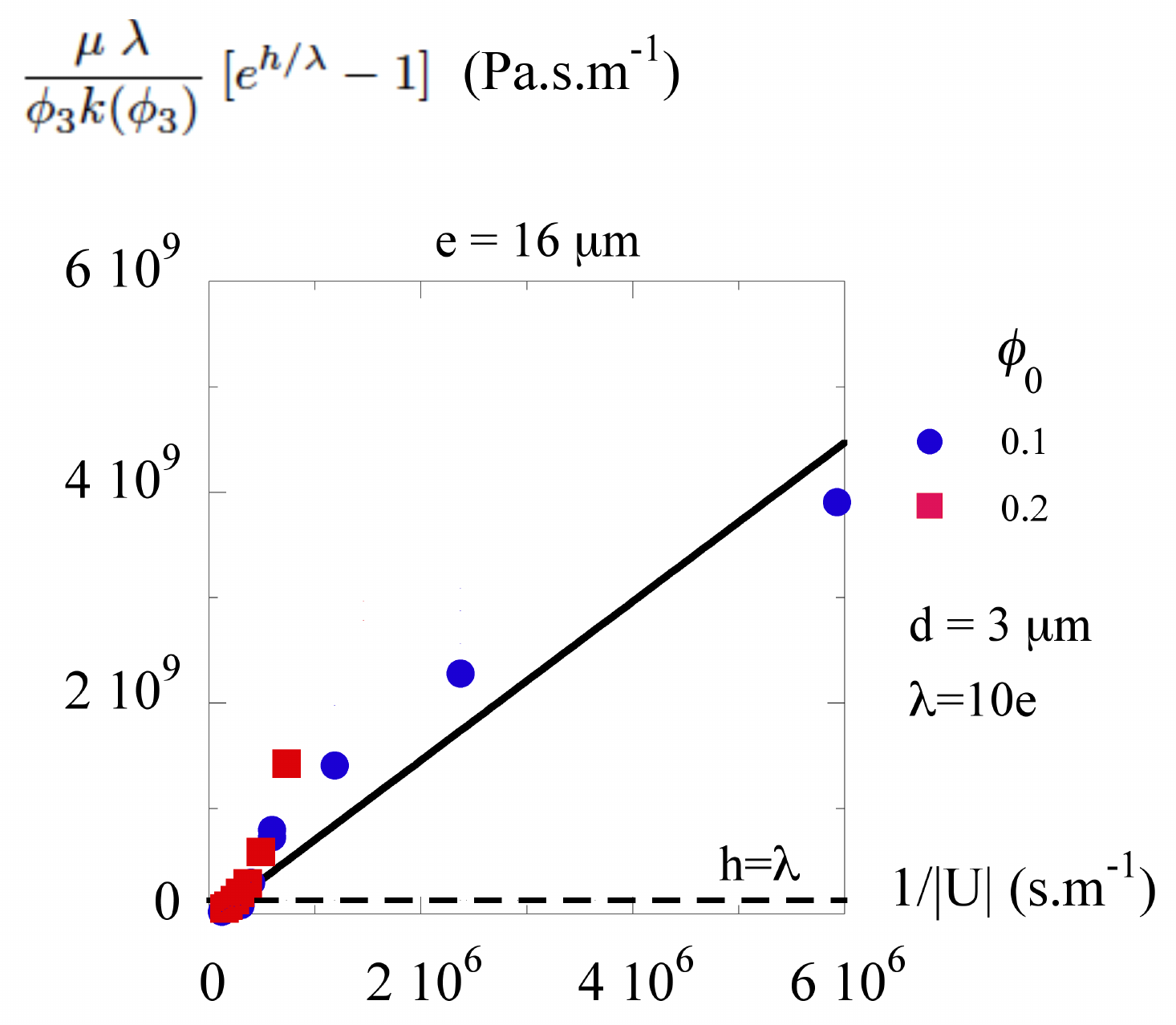}
\hspace{1cm}
\includegraphics[height=6cm]{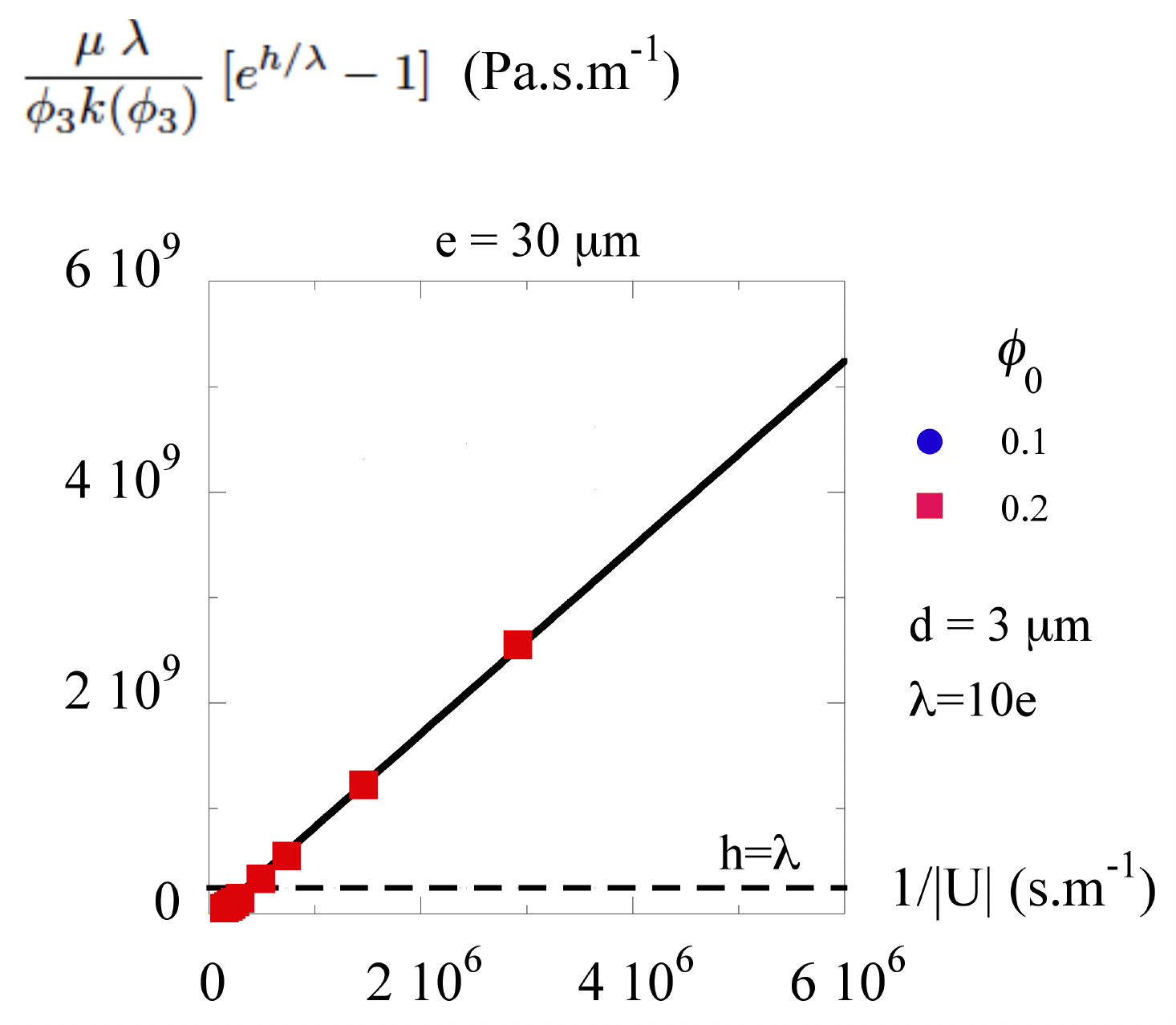}
\\
	\flushleft{\hspace{4cm} (a) \hspace{8cm} (b)}
\centering
\includegraphics[height=6cm]{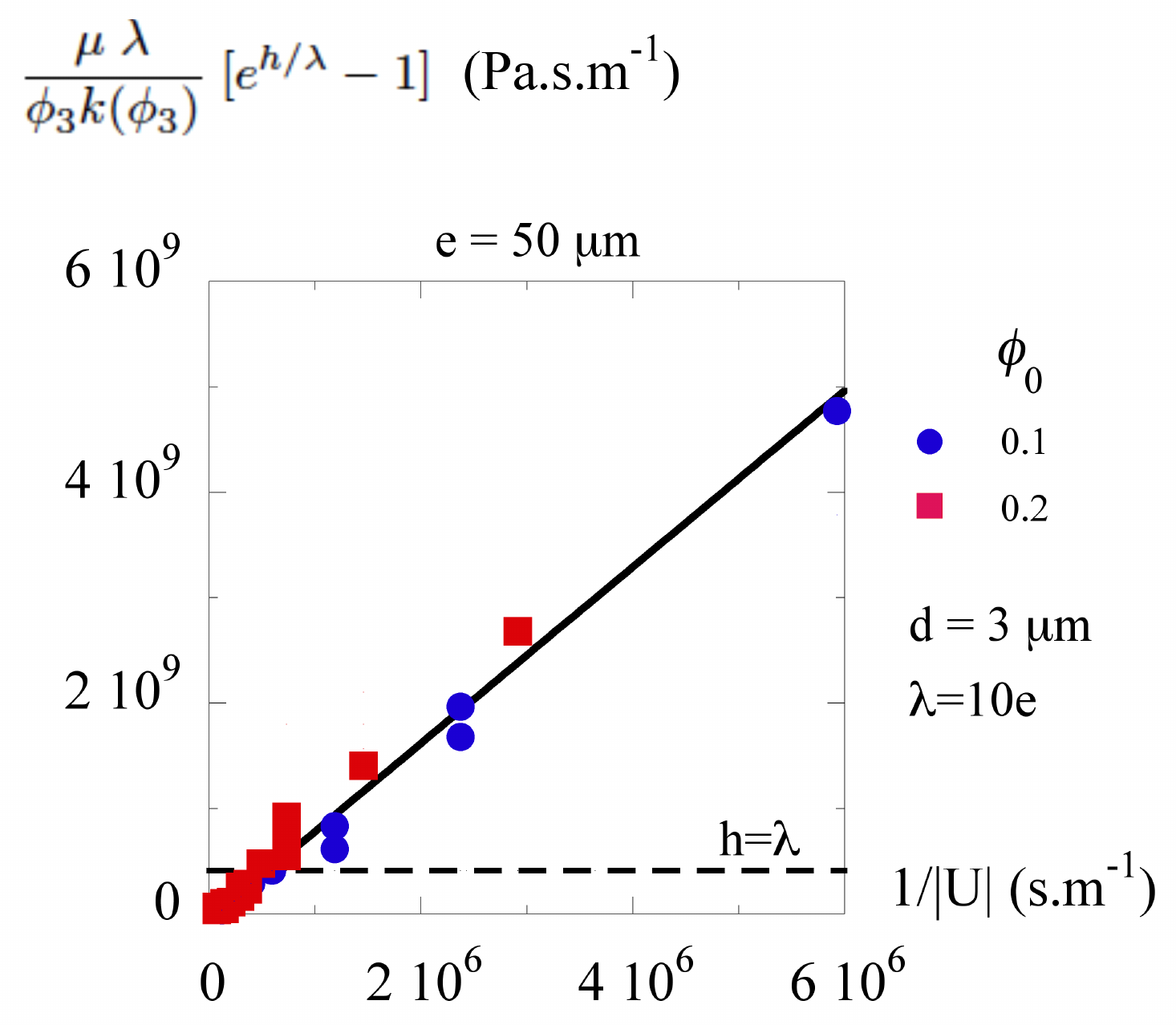}
\hspace{1cm}
\includegraphics[height=6cm]{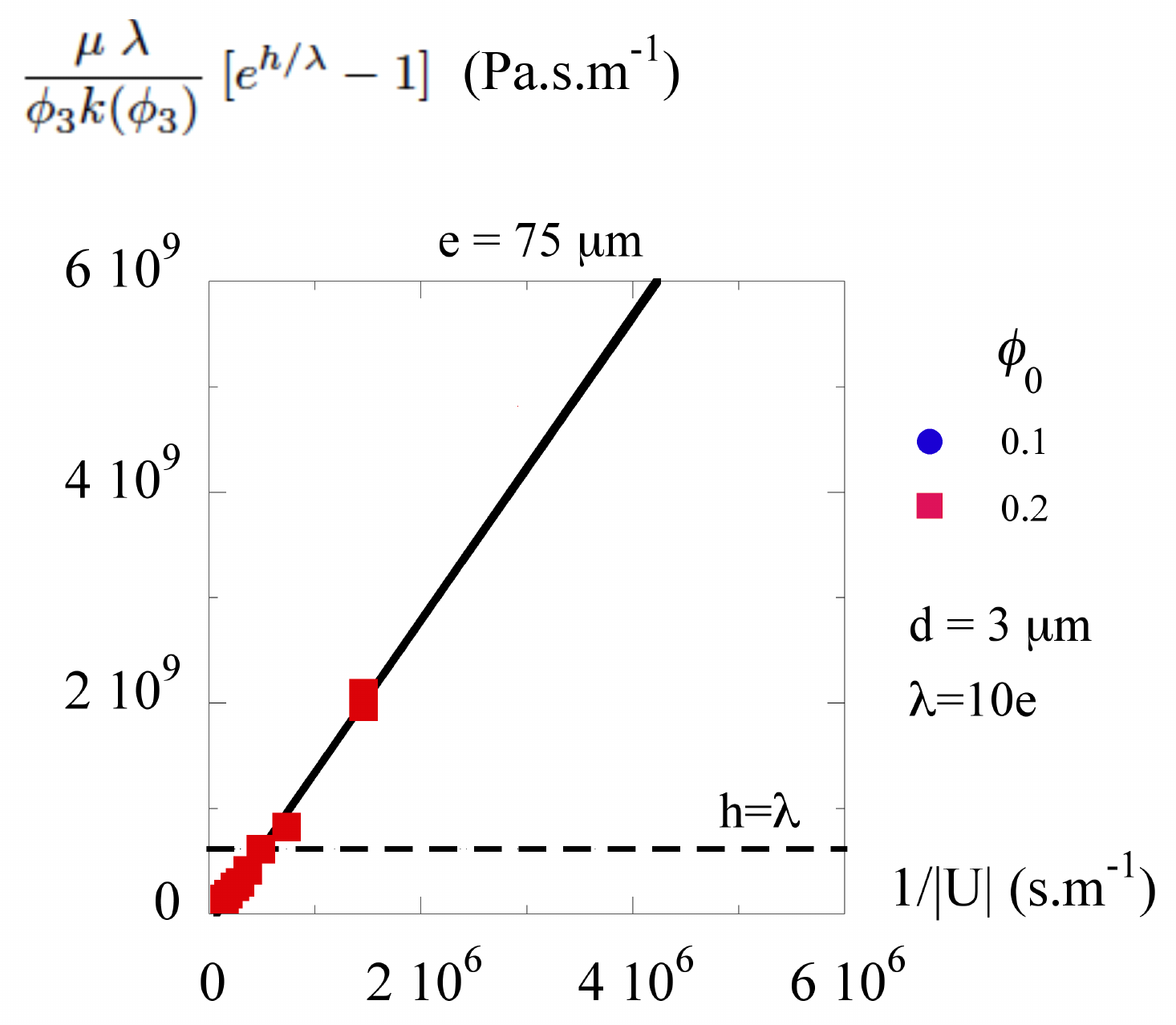}
\\
	\flushleft{\hspace{4cm} (c) \hspace{8cm} (d)}
\centering
\includegraphics[height=6cm]{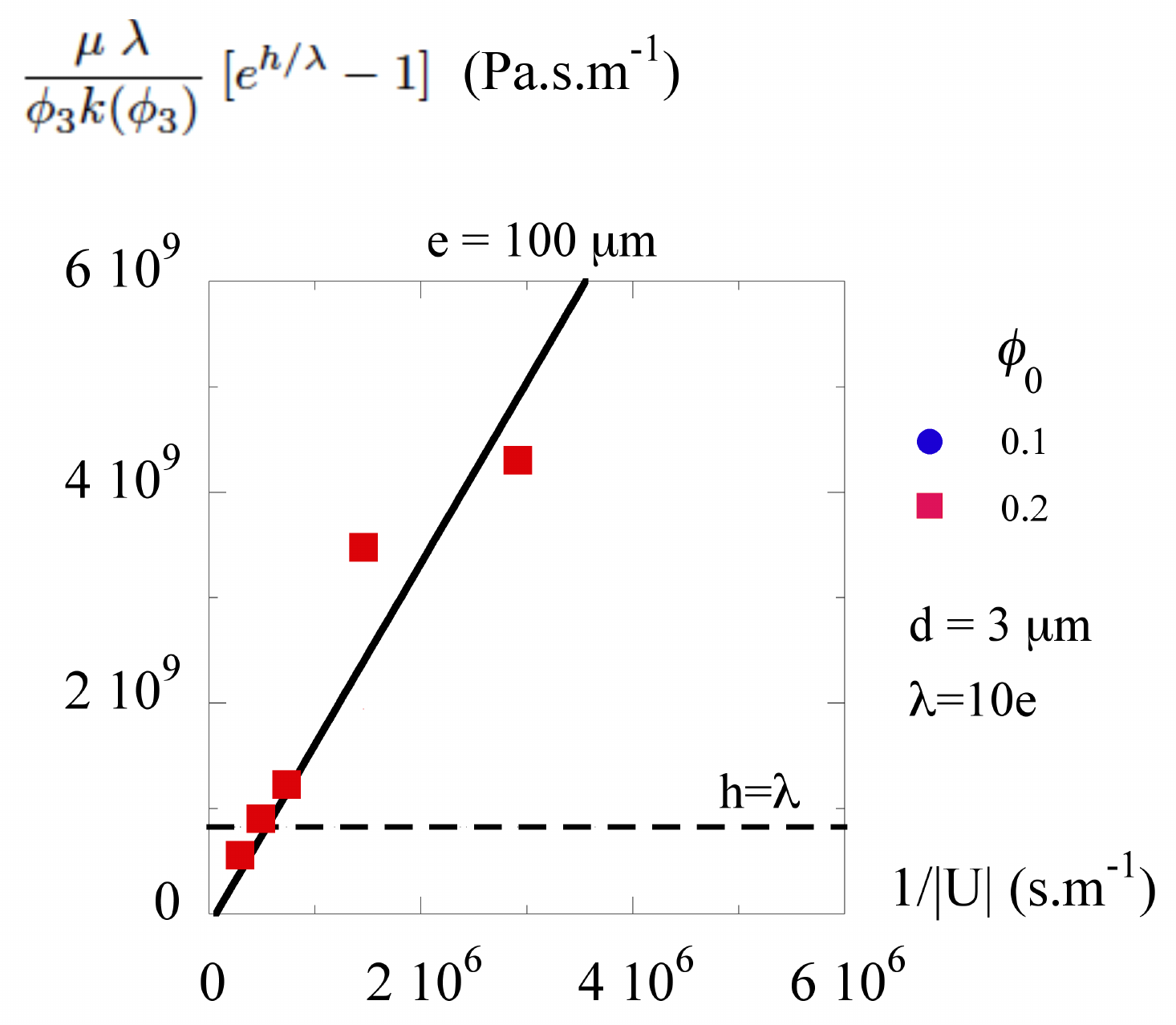}
\hspace{1cm}
\includegraphics[height=6cm]{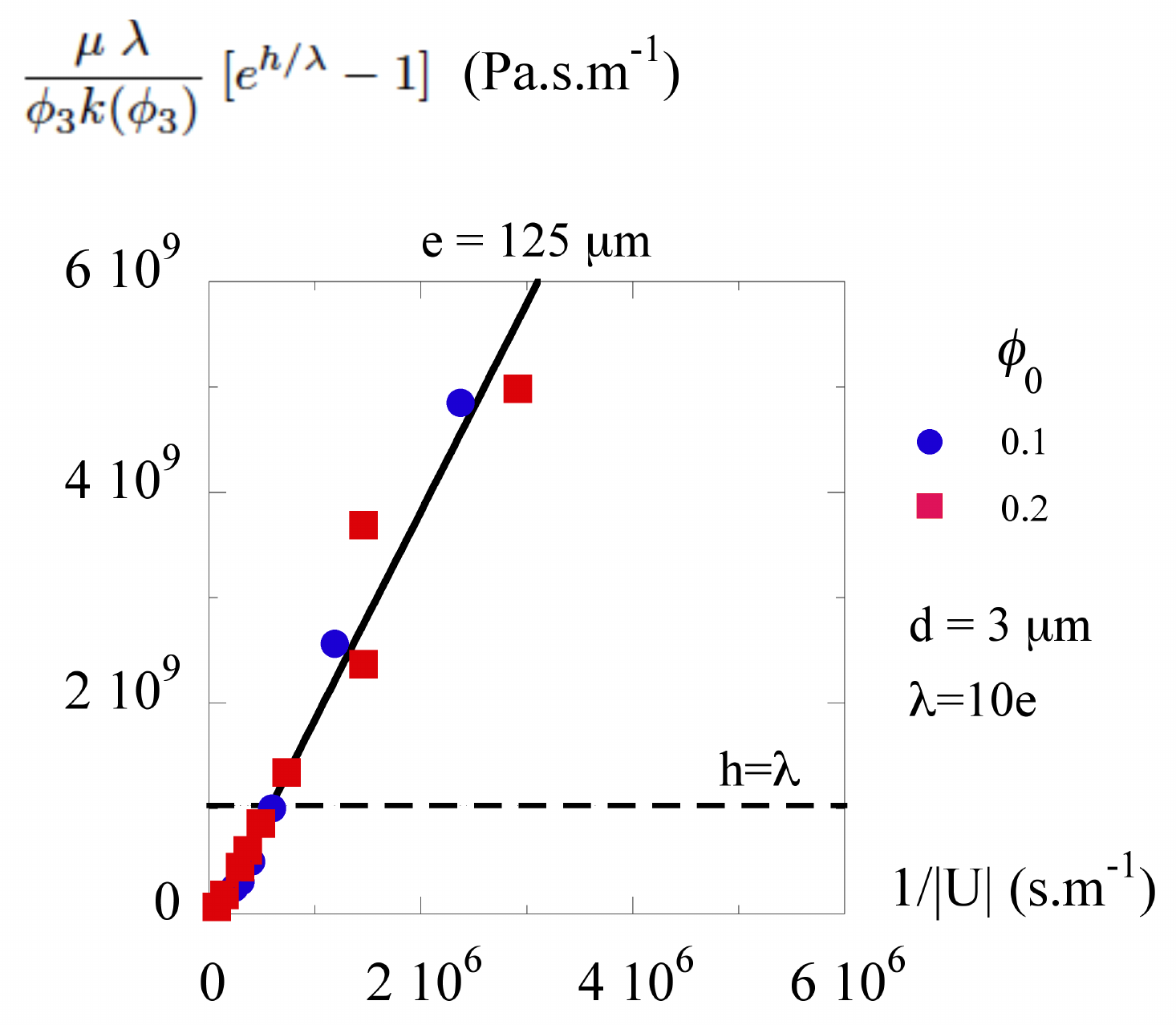}
\\
	\flushleft{\hspace{4cm} (e) \hspace{8cm} (f)}
	\vspace{1cm}
\caption{(Color online)
Evolution of the combination $\lambda \; [\exp(\frac{h}{\lambda}) - 1]$ of the compacted layer thickness $h$ with the inverse of the Darcy velocity $U$ for various sample depths $e$.
The value of $\lambda$ is fixed at $\lambda=10.0 e$.
Following relation (\ref{h,1/U,1/Uc}), in order to provide a measure of $\bar{P}_T$ from its slope, the exponential combination is multiplied by the factor $\mu/[\phi_3 k(\phi_3)]$ where $\phi_3$ denotes the particle volume fraction at random close packing and $k(\phi_3)$ the resulting permeability at the actual particle diameter $d=3\mu$m.
Thick lines show the resulting linear fits with  $U_c$ fixed at $15 \mu$m$.s^{-1}$.
Dashed lines show the values $h=\lambda$ in the units of the ordinates.
Sample depths : (a) $e=16 \mu$m. (b) $e=30 \mu$m.
(c) $e=50 \mu$m. (d) $e=75 \mu$m.
(e) $e=100 \mu$m. (f) $e=125 \mu$m.
}
\label{G,PT,U,e}
\end{figure*}

\begin{figure}[t!]
\centering
\includegraphics[height=6cm]{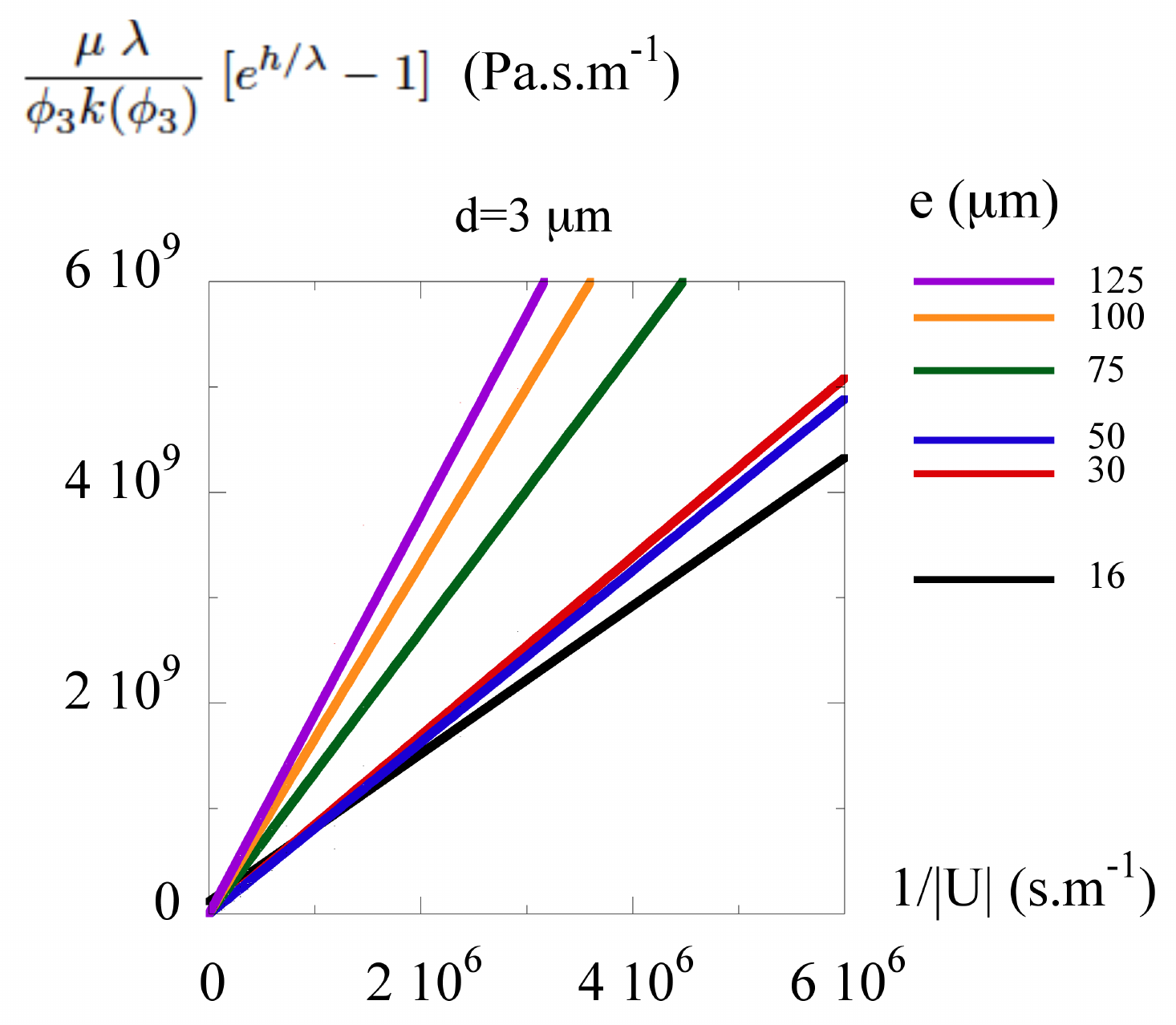}
\\
\caption{(Color online)
Overview of the evolutions of the combination $\lambda \; [\exp(\frac{h}{\lambda}) - 1]$ with the inverse of the Darcy velocity $U$ for different sample depths $e$.
The lines correspond to the fits of the linear evolutions evidenced in figure \ref{G,PT,U,e} for sample depths $e$ varying from $16 \mu$m to $125 \mu$m.
They show a slope, and thus a thermomolecular pressure $\bar{P}_T$, that increases with the sample depth.
}
\label{G,PT,U,all}
\end{figure}

\section{Particle layer inhomogeneity}
\label{ParticleLayerInhomogeneity}

The dependence of the layer thickness $h$ on the sample depth $e$ suggests an effect of the boundaries, here the sample plates.
It cannot be hydrodynamical since the compactness of the layer restricts the hydrodynamic boundary length to less than a particle diameter, a quantity too small to induce a global effect on the system.
On the other hand, the flatness of the plates introduces a long-range correlation of position that can significantly affect the distribution of particles, and thus the physical features of the compacted layer.
We address below this effect and its implication on the layer thickness.

\subsection{Sample plate and particle ordering}
\label{PlateOrdering}

\begin{figure*}[t!]
\centering
\includegraphics[height=5cm]{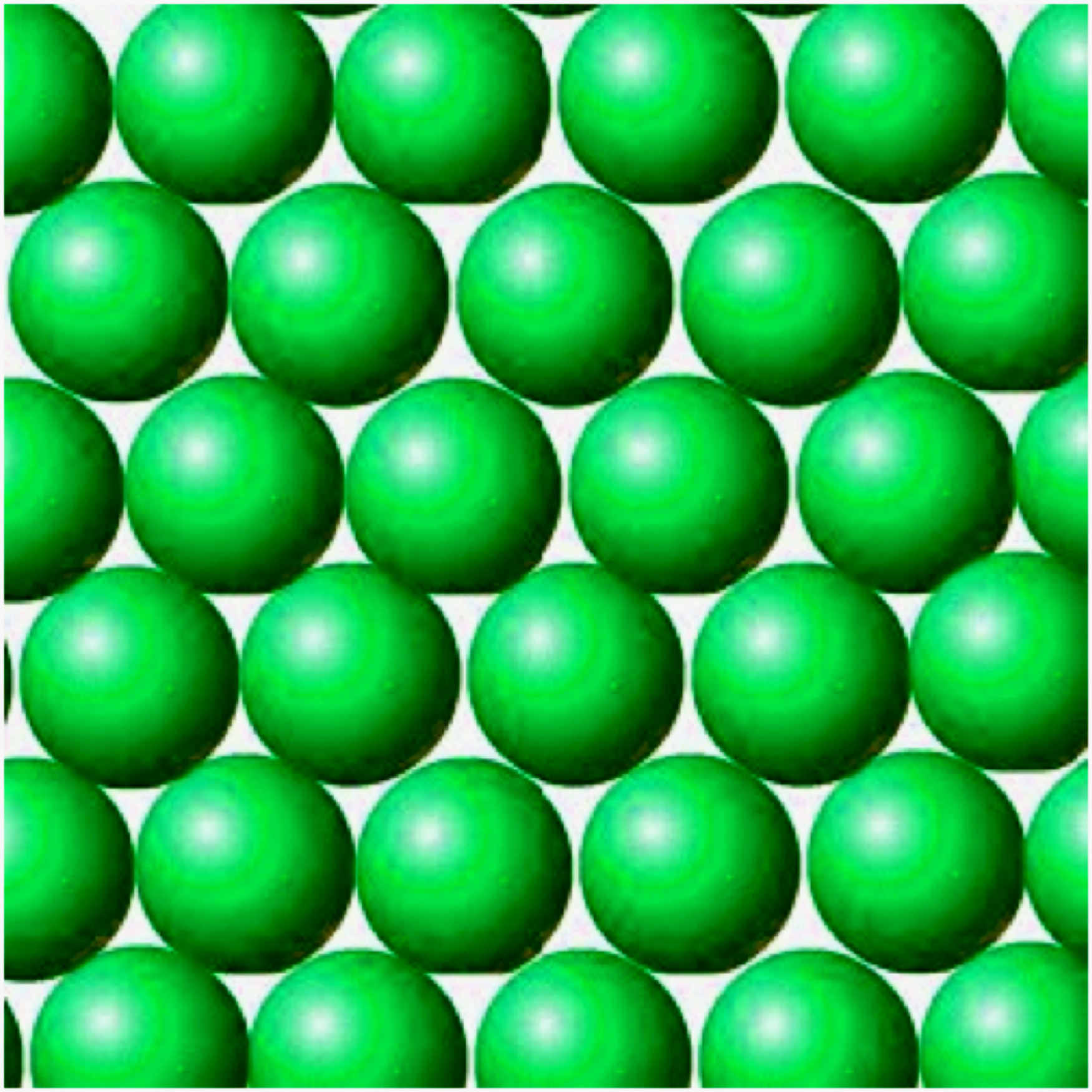}
\hspace{0.2cm}
\includegraphics[height=5cm]{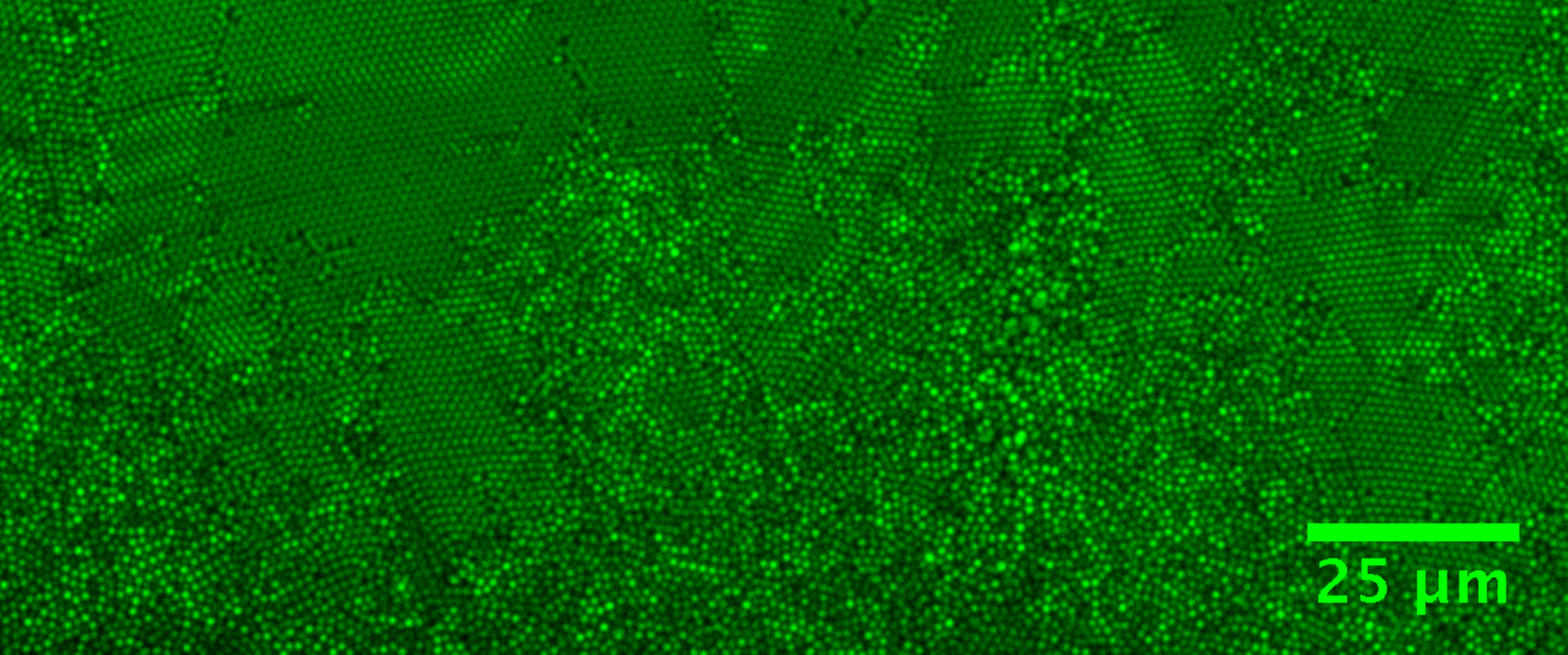}
\caption{(Color online)
(a) Hexagonal close-packed lattice of spheres. 
This corresponds to the arrangement of spheres of highest density : $\phi_2= \pi/3\sqrt{2} \approx 0.74048$.
It is compatible with a planar boundary and is thus the highest density available at a sample plate.
This density can prolongate away from the boundary with an hexagonal close-packed or a face-centered cubic lattice.
(b) Confocal microscopy image of the compacted layer of particles close to a sample plate. 
The particle diameter is $1~\mu$m. 
The image shows large close-packed arrangements of particles with hexagonal order in a plane parallel to the plate.
This implies a particle volume fraction close to $\phi_2$ at the sample plates.
}
\label{Hexagonal}
\label{Confocal}
\end{figure*}

In 1611, Kepler conjectured that the highest density of sphere packing in space, whatever the regular or irregular nature of its arrangement, was achieved with either an hexagonal close-packed or a face-centered cubic lattice.
These lattices are made of superposed planar hexagonal arrangement [Fig. \ref{Hexagonal}(a)] with different phases of superposition.
In 1831, Gauss proved this conjecture in the restricted case of regular lattices.
In the general case of arbitrary arrangements, it was finally proven by Hales  \cite{Hales2005} 20 years ago and later certified by a formal proof.
The density of these hexagonal close-packed arrangements amounts to $\phi_{hcp}=\pi/3 \sqrt{2} \approx 0.74048$.

As these highest density arrangements of spheres are made of superposed planar lattices, they fit with a planar boundary.
We shall call hereafter their density $\phi_2$ to state that it corresponds to the highest density achievable in a \emph{plane} : $\phi_2=\phi_{hcp}$.

When random arrangements of particles in space are considered instead of lattices, the highest possible density is smaller.
It has been found experimentally to be $\phi_{rcp}=0.6366\pm0.0005$ \cite{Scott1969} and has been shown to be at most $0.634$ by statistical analysis of jammed states \cite{Song2008}, although the concept of random arrangements may require clarification \cite{Torquado2000}.
We retain the latter determination for the highest density of random close packing.
We recall that, in section \ref{Particle layer thickness}, we have labeled its density $\phi_3$ to state that it refers to random sphere arrangements in \emph{space}.

Regarding the compacted layers of particles, as they are built by piling up incoming particles, one may expect them to correspond to the highest density compatible with this process.
Following the spatial constraint implied by the sample plates, one has however to distinguish their bulk from their plate vicinity.

In the bulk, as the packing is random, the random close packing density $\phi_{rcp}=\phi_3$ is expected.
It has been actually evidenced as the mean particle volume fraction at the largest sample depth $e=125 \mu$m (Fig. \ref{DST}) \cite{Saint-Michel2017}.
As the boundary effects of samples plates are minimized at this large depth, this supports a packing volume fraction equal to $\phi_{3}$ in the layer bulk.
  
At a sample plate, impenetrability of this flat boundary forces particles  to align on its planar surface. 
This geometrical constraint breaks the 3D random packing and induces correlations in particles positions, so that modifications of the particle volume fraction $\phi_l$ are expected.
In particular, in granular materials, a wall-induced ordering has been largely evidenced and documented in monodisperse \cite{Zhang2006,Burtseva2015,Mandal2017} or bidisperse mixtures \cite{Desmond2009}.
It may even induce crystallization near the walls as found for monodisperse mixtures in channel flows  \cite{Mandal2017}, Couette flows \cite{Mueth2003}, or by shaking \cite{Pouliquen1997}.
 
Here, confocal microscopy evidences in figure \ref{Confocal}(b) a particle arrangement at the sample plates made of hexagonal patches linked by crystalline defects.
This corresponds to a  crystallization presumably favored by the pressure exerted by the particle matrix on the particles nearing the sample plates.
As lines of defects negligibly modify the particle density, the particle volume fraction at the sample plate is thus close to that of an hexagonal array, $\phi_2$.

Our observations and measurements thus evidence two different particle volume fractions, $\phi_{3}$ in the bulk and $\phi_{2}$ at the sample plates.
In between, the particle volume fraction increases from $\phi_{3}$ to $\phi_{2}$ as the sample plate is approached (Fig.\ref{I,Layer}).
We address below the mechanical implications of this layer heterogeneity and propose some models for the transition of the particle density from $\phi_2$ to $\phi_3$.

\begin{figure}[t!]
\centering
\includegraphics[height=4cm]{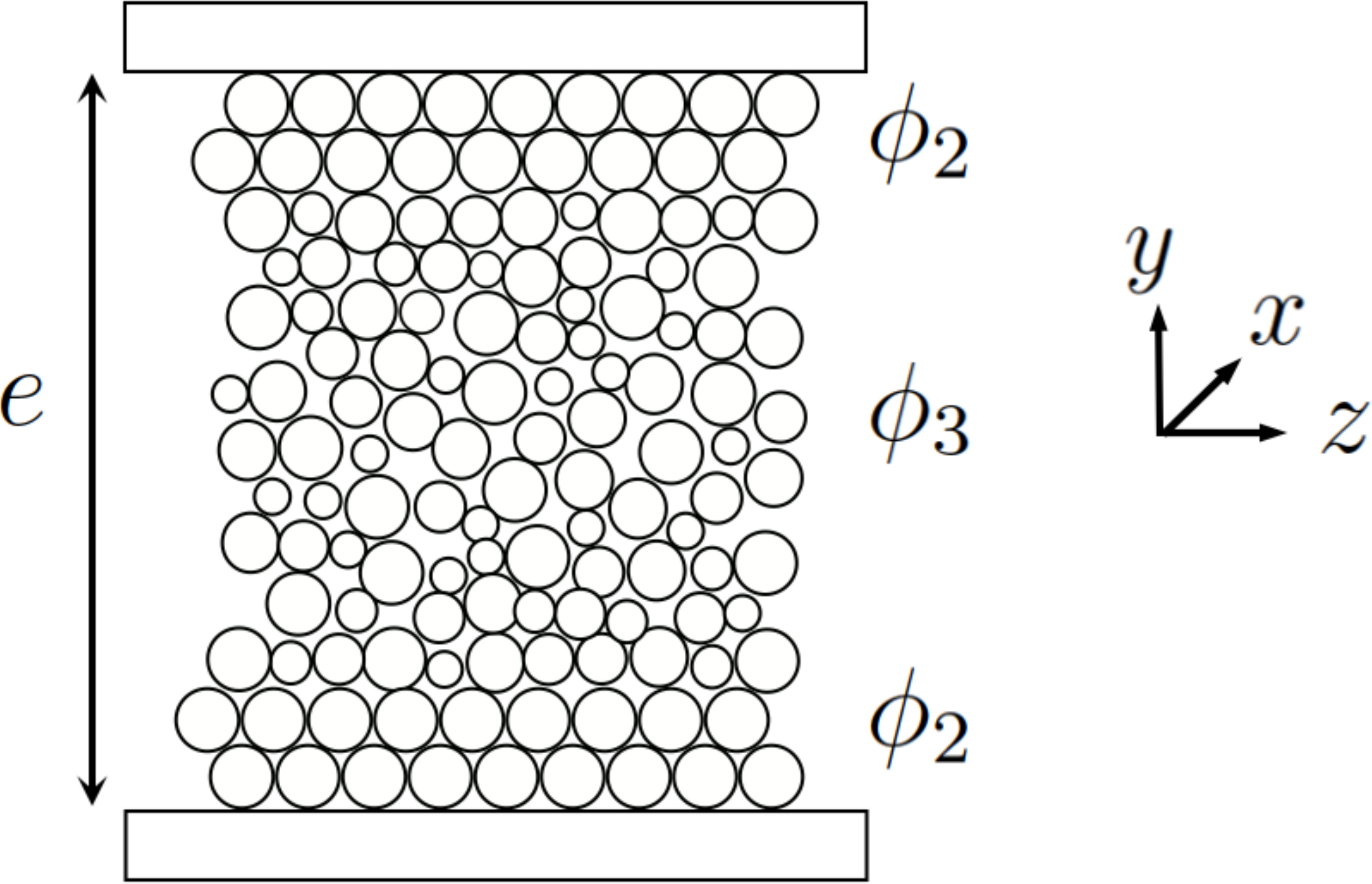}
\\
\caption{(Color online)
Sketch of a cross-section of the compacted layer showing and ordered arrangement of particles near the plates (density $\phi_2=\phi_{hcp}$) and a random one far from them (density $\phi_3=\phi_{rcp}$). 
Variations of particle diameters is an artifact of the cross-section on the random part of the layer.
}
\label{I,Layer}
\end{figure}

\subsection{Implication on the particle matrix pressure}
\label{Modeling}

The implication of particle volume fraction on the matrix pressure comes from the Darcy velocity (\ref{U}), the permeability (\ref{KC}) and the normalization of pressure (\ref{NormalizationPressure}).
However, among these three factors, the most important is the permeability $k(\phi_l,d)$ which, owing to the term $(1-\phi_l)^3$, decreases by a factor nearly 4 as $\phi_l$ rises from $\phi_3$ to $\phi_2$.
We thus call this interplay between particle volume fraction and matrix pressure the "permeability mechanism".
We quantify it below at the largest and thinnest depths and then extrapolate it to intermediate depths.

From relations (\ref{U}) (\ref{KC}) and (\ref{Pp,U}), the pressure exerted by a particle matrix of density $\phi_l$ on particles nearing the front reads as :
\begin{equation}
\label{Pp,V}
\bar{P_{\rm p}} = -  f(\phi_l,\phi_0) \; V \; \lambda \; [\exp(h/\lambda)-1]
\end{equation}
with a dissipation factor $f$ :
\begin{equation}
\label{f}
f(\phi_l,\phi_0) =\mu \frac{180}{d^2}\frac{(\phi_l-\phi_0)}{(1-\phi_l)^3}.
\end{equation}

This makes the force balance (\ref{h,1/U,1/Uc}) expresses as :
\begin{equation}
\label{PT,h,1/V,1/Vc,f}
\bar{P_{\rm T}} \, (\frac{1}{V} - \frac{1}{V_c}) =  f(\phi_l,\phi_0)\; \lambda \; [\exp(\frac{h}{\lambda}) - 1],
\end{equation}
where $V_c$ denotes the velocity corresponding to $U_c$ at $\phi_l$, according to (\ref{U}).
Therefore, for given data $(V, h)$, an error in the evaluation of $\phi_l$ would result from $f(\phi_l,\phi_0)$ in some error in the estimation of $\bar{P_{\rm T}}$.

To test whether the permeability mechanism may solely explain the variations of $\bar{P_{\rm T}}$ this way, we consider the largest and smallest sample depths, $e=125$ and $16 \mu$m, and assume their particle layers to be homogeneous with particle densities $\phi_3$ and $\phi_2$ respectively.
This approximation turns out neglecting boundary effects at the largest depth since most of the particles stand far from the plates and, at the smallest depth, the relaxation to the bulk value since all particles stand at a distance $e/2=8 \mu$m less than three particle diameters from the plates.

Then, considering as in section \ref{Thermomolecular pressure} that $\phi_l=\phi_3$  in the whole sample whatever the sample depth, would provide the correct value for $e=125 \mu$m but would underestimate for $e=16 \mu$m both $f$ and $\bar{P}_T$  by a factor $\alpha = f(\phi_2,\phi_0)/f(\phi_3,\phi_0)\approx 3.4$.
Interestingly, this factor is of the same order than the ratio $2.62$ between the values of $\bar{P}_T(e)$ at $e=125 \mu$m and $e=16 \mu$m deduced from figure \ref{G,PT,U,e}.
As the thermomolecular pressure $\bar{P}_T$ is expected to be independent of the sample depth $e$, these close values provide a strong support to the implication of the particle volume fraction on the apparent variation of $\bar{P}_T$ with $e$ and in particular to the relevance of the permeability mechanism.

However, the above statement has been established for homogeneous volume fraction $\phi_l$, either $\phi_2$ or $\phi_3$, whereas $\phi_l$ actually varies in the layer from the value $\phi_2$ close to the plates to the value $\phi_3$ far from them (Fig. \ref{I,Layer}).
Taking into account this transition qualitatively explains the continuous increase with $e$ of the slopes (and thus of $\bar{P}_T$) in figures \ref{G,PT,U,e} and \ref{G,PT,U,all}.
Conversely, this means that the relation $\bar{P}_T(e)$ determined in section \ref{Thermomolecular pressure} encodes the way the volume fraction $\phi(y)$ changes from $\phi(0)=\phi_2$ at the plates to $\phi(\infty)=\phi_3$ far from them.
For instance, one may expect that the stagnation of $\bar{P}_T(e)$ between $e=16 \mu$m and $e=50 \mu$m displayed in figure \ref{G,PT,U,all} goes together with a stagnation of $\phi(y)$ for a distance between $8$ and $25 \mu$m from the plates, before decreasing to $\phi_3$.

To provide insights into the evolution of $\phi(y)$, we thus propose to consider different relevant modelings of $\phi(y)$, determine their resulting evolutions $\bar{P}_T(e)$ (\ref{PT,h,1/V,1/Vc,f}) according to the permeability mechanism and finally compare them to that deduced from the data of figures \ref{G,PT,U,e} and \ref{G,PT,U,all}.
This way, we expect first to evidence the features of the particle layer inhomogeneity $\phi(y)$ that play a major role in the difference between the suspension behaviors observed in figure \ref{G,PT,U,all} and, second, to determine the resulting common value of the thermomolecular pressure $\bar{P_{\rm T}}$.

For this we first notice that, from relation (\ref{PT,h,1/V,1/Vc,f}), a unique $\bar{P_{\rm T}}$ with a variable $\phi_l(y)$ would yield a layer thickness $h$ varying with $y$ from one plate to the other.
However, the images of the compacted layer obtained both by transmission and reflection (Fig. \ref{ReflectionTransmission}) show that the layer thickness is actually the same close to the plates and far from them.
This paradox turns back to the expression(\ref{Pp,V}) of the particle matrix pressure $\bar{P_{\rm p}}$ following which, for a constant $h$ and a variable $\phi_l(y)$, the pressure $\bar{P_{\rm p}}$ should vary from one plate to the other.
This is what would actually happen if the columns of particles that stand at a given distance $y$ from the plates involved no mechanical exchanges.
However, it is known that stresses in the particle matrix redistribute homogeneously as described by the Janssen's redirection coefficient \cite{Janssen1895,Bertho2003,AndreottiForterrePouliquen2013}.
In addition, fluid pressure tends also to equilibrate by fluid motion.
These mechanisms thus make the particle pressure $\bar{P_{\rm p}}$ on the particles nearing the solidification front homogeneous.
Accordingly, we shall express $\bar{P_{\rm p}}$ as the average over the sample depth of the pressures $\bar{P_{\rm p}}(y)$ determined form (\ref{Pp,V}).
We shall denote this common pressure  $\bar{P_{\rm p}}^i$, the superscript recalling that it refers to an inhomogeneous distribution of particle volume fraction :
\begin{equation}
\label{Pmui}
\bar{P_{\rm p}}^i = - \frac{1}{e} \int_{-e/2}^{e/2}f(\phi,\phi_0) dy \; V \;  \lambda \; [ \exp(\frac{h}{\lambda}) -1]
\end{equation}
with an origin of the y-axis placed in the middle of the layer depth (Fig. \ref{I,Layer}) and a volume fraction $\phi$ \emph{a priori} dependent on $y$ : $\phi\equiv \phi(y)$.

In contrast, we shall label with a superscript $h$ the previous expression of $\bar{P_{\rm p}}$ (\ref{Pp,V}) that has been used in figures (\ref{G,PT,U,e}) (\ref{G,PT,U,all}) for a homogeneous layer with particle volume fraction $\phi_l = \phi_3$.
Both determinations are connected by :
\begin{equation}
\label{Pmu,i,h}
\bar{P_{\rm p}}^i =  I\bar{P_{\rm p}}^h,
\end{equation}
where $I\equiv I[\phi(y),\phi_0,e]$ denotes the renormalization factor :
\begin{equation}
\label{I}
I = \frac{1}{e} \int_{-e/2}^{e/2} \frac{f(\phi,\phi_0)}{f(\phi_3,\phi_0)} dy
\end{equation}
which conveys the effects of inhomogeneity.

On the other hand, we note from (\ref{U}) that the relative variations of $U$ due to $\phi_l$ varying from $\phi_2$ to $\phi_3$ are less than $8\%$ whereas the ratio $U/U_C$ is smaller than $0.5$ on our data base. 
Altogether, in the force balance (\ref{ForcebalancePpPT}), this makes the factor $(1-|U|/U_c)$ vary by less than $8\%$ so that the change of thermomolecular pressure $\bar{P_{\rm T}}$ implied by layer inhomogeneity follows that found on $\bar{P_{\rm p}}$.
Calling $\bar{P_{\rm T}}^h$ the previous evaluation of $\bar{P_{\rm T}}$ on an homogeneous layer at $\phi_l=\phi_3$ and $\bar{P_{\rm T}}^i$ its value on inhomogeneous layers, one then obtains :
\begin{equation}
\label{PT,i,h}
\bar{P_{\rm T}}^i = I \; \bar{P_{\rm T}}^h.
\end{equation}

Our ansatz is thus that  the dependence $\bar{P_{\rm T}}^h(e)$ on the sample depth $e$ results from an incorrect assumption of uniform particle volume fraction $\phi_l=\phi_3$ in the compacted layer and that it could be corrected by considering an appropriate distribution $\phi(y)$ ranging from $\phi_2$ at the sample plates to $\phi_3$ at the bulk.
Then, applying the corresponding renormalization factor $I$ should yield the correct value $\bar{P_{\rm T}}^i$ to be recovered.
A test of the relevance of this ansatz and of the resulting value of $\bar{P_{\rm T}}^i$ will be its independence on $e$.

Our objective will now be to uncover what suitable distribution of particle volume fraction $\phi(y)$ could recover the dependence of $\bar{P_{\rm T}}^h$ with $e$ and provide the actual $e$-independent thermomolecular pressure $\bar{P_{\rm T}}^i$.

\subsection{Volume fraction models}
\label{VolumeFractionModels}

By symmetry, we expect an even distribution $\phi(y)$ of particle volume fraction with respect to the mid-plane $y=e/2$ (Fig. \ref{I,Layer}).
This will enable us to confine the computation of the inhomogeneity factor $I$ to a half layer.
Then, for convenience, we change the origin of the $y$-axis and place it on the bottom plate.
This way the plate location, $y=0$, is fixed and that of the mid-layer, $y=e/2$, evolves with the layer thickness.

We shall first consider a sharp transition between hexagonal and random close packing and then an exponential relaxation between them.
It will finally appear that a combination of both will be required to recover the variations observed with the sample depth $e$.

\begin{figure*}[t!]
\centering
\includegraphics[width=8cm]{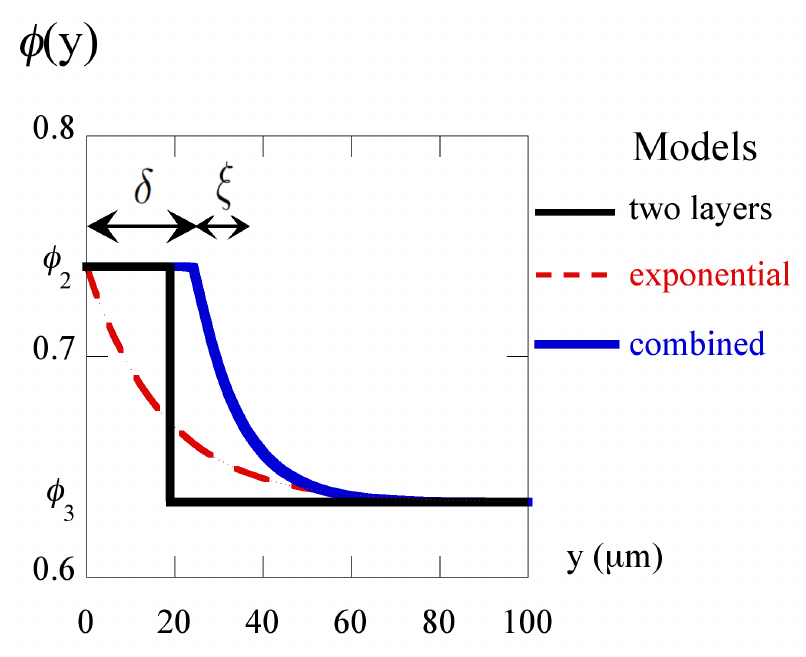}
\hspace{1cm}
\includegraphics[width=8cm]{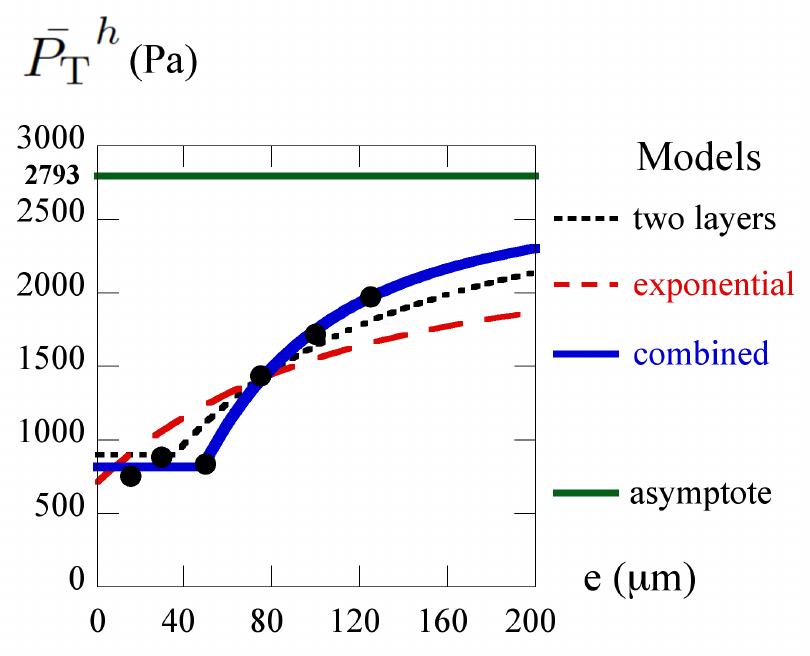}
\caption{(Color online)
Models and implications. 
(a) Models of the evolution $\phi(y)$ of the particle volume fraction from the sample plate $y=0$.
The lengths $\delta$ and $\xi$ correspond to the depth of the homogeneous crystallized layer at the plate and to the relaxation length towards the bulk value $\phi_3$.
(b) Best fits of the resulting evolution of the thermomolecular pressure $\bar{P_{\rm T}}^h$ with the sample depth $e$. 
Points correspond to the measures of $\bar{P_{\rm T}}^h$ as provided by the slopes of figures \ref {G,PT,U,e} and \ref{G,PT,U,all}.
The green line indicates the asymptote of the combined model.
}
\label{G,phi,y}
\label{G,PT,e,Lin,Alle,fits}
\end{figure*}

\begin{figure}[t]
\centering
\includegraphics[width=8cm]{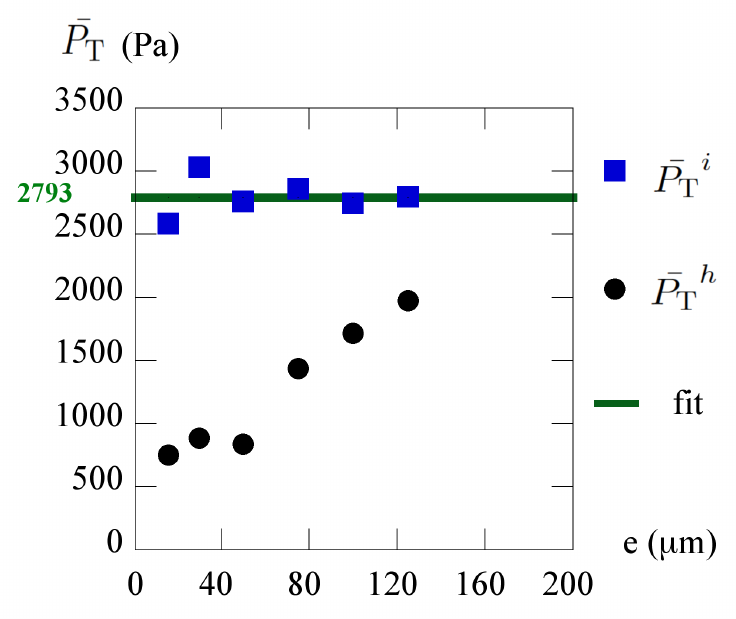}
\caption{(Color online)
Renormalization of the thermomolecular pressure $\bar{P}_T$ on particles entering the front following the combined model.
Full circles : pressures $\bar{P_{\rm T}}^h (e)$ determined for homogeneous layers.
Full squares : pressures $\bar{P_{\rm T}}^i (e)$ determined for inhomogeneous layers.
Full line : best fitted value of $\bar{P_{\rm T}}^i$ in the combined model of figure \ref{G,PT,e,Lin,Alle,fits}(b).
}
\label{G,PT,PTe}
\end{figure}

\subsubsection{Two layers model}
\label{Two layers model}
We consider the compacted half-layer as composed of two homogeneous sub-layers.
One, of depth $\delta$, is adjacent to the sample plate.
It thus involves the highest particle density $\phi_2$.
The other extends up to the mid-layer and therefore involves the random close packing density $\phi_3$.

This model corresponds to a sharp transition between the limit densities $\phi(0)=\phi_2$ and $\phi(\infty)=\phi_3$ :
\begin{eqnarray}
\nonumber
0 \leq y \leq \delta &:& \phi(y)=\phi_2, \\
\nonumber
\delta  \leq y  \leq e/2 &:& \phi(y)=\phi_3.
\end{eqnarray}

We label $\alpha(\phi_0)=f(\phi_2,\phi_0) / f(\phi_3,\phi_0)$ the increase of dissipation factor $f$ when changing the volume fraction from $\phi_3$ to $\phi_2$.
Its values at $\phi=0.1$ and $0.2$ are close, respectively $3.364$ and $3.493$.
We denote $\alpha$ their average value $3.429$ and, for simplicity, use it as representative of the two volume fractions studied.

We then obtain from (\ref{I}) the inhomogeneity factor :
\begin{eqnarray}
\nonumber
0 \leq e \leq 2 \delta &:& I(\delta,e)=\alpha, \\
\nonumber
2\delta  \leq e &:& I(\delta,e)=1 + (\alpha-1) 2 \delta / e.
\end{eqnarray}

The values of $\bar{P_{\rm T}}^h(e)$ obtained from the slopes of figure \ref{G,PT,U,all} are reproduced on figure \ref{G,PT,e,Lin,Alle,fits}(b).
The best fit of $\bar{P_{\rm T}}^h (e) = I(\delta,e)^{-1} \bar{P_{\rm T}}^i$ with fitting parameters $\delta$ and $\bar{P_{\rm T}}^i$ yields $\delta = 18.1 \mu$m and $\bar{P_{\rm T}}^i = 3082 Pa$.
The corresponding profiles of $\phi(y)$ and of $\bar{P_{\rm T}}^h (e)$ for this so-called two layers model are reported on figures \ref{G,PT,e,Lin,Alle,fits} (a) and (b).
It appears that both the stagnation at low $e$ and the rise beyond are are too weak.

\subsubsection{Exponential relaxation model}
\label{Exponential}

As a sharp transition between two packing densities is crude, we wish to model a continuous transition between them.
We then consider an exponential relaxation of $\phi(y)$ from the value $\phi_2$ at the sample plate to the value $\phi_3$ far from it :
\begin{equation}
\label{phi, exp}
\phi(y) = \phi_3 + (\phi_2- \phi_3) \exp(-y/\xi),
\end{equation}
where $\xi$ denotes a relaxation length.

Standard integration from relations (\ref{I}) and (\ref{phi, exp}) yields the inhomogeneity factor $I(\xi,e)$ :
\begin{eqnarray}
\label{I,exp}
\nonumber
&&I(\xi,e) = 1 + \frac{1}{\epsilon} \big\{- \ln(1-\beta) + (1-\beta)^{-1} + \frac{\gamma}{2} (1-\beta)^{-2} \\
&&+ \ln (1-\beta e^{-\epsilon}) - (1-\beta e^{-\epsilon})^{-1} - \frac{\gamma}{2} (1-\beta e^{-\epsilon})^{-2} \; \big\}
\end{eqnarray}
with :
\begin{equation}
\label{beta,gamma,espilon}
\beta=\frac{\phi_2 -\phi_3}{1-\phi_3} \; ; \;  \gamma=\frac{1-\phi_0}{\phi_3-\phi_0} \; ; \;  \epsilon=\frac{e}{2\xi}.
\end{equation}

Note that $I$ involves the same limits as in the two layers model since $I(\xi,0)=\alpha$ and $I(\xi,\infty)=1$.
The explicit values are $\beta = 0.291$, $\gamma=1.685$ for $\phi_0=0.1$ and $\gamma=1.843$ for $\phi_0=0.2$.
Following the close values of $\gamma$, we adopt for simplicity its average value $1.764$ for the two volume fractions studied.

The best fit of $\bar{P_{\rm T}}^h (e) = I(\xi,e)^{-1} \bar{P_{\rm T}}^i$ with fitting parameters $\xi$ and $\bar{P_{\rm T}}^i$ yields $\xi = 17.5 \mu$m and $\bar{P_{\rm T}}^i = 2397$ Pa.
The corresponding profiles of $\phi(y)$ and of $\bar{P_{\rm T}}^h (e)$ for this so-called exponential model are reported on figures \ref{G,PT,e,Lin,Alle,fits} (a) and (b).
It appears that neither the stagnation at low $e$ nor the magnitude of the rise above are correctly reproduced.

\subsubsection{Combined model}
\label{Combined}

Figure \ref{G,PT,e,Lin,Alle,fits}(b) reveals that the above models fail in recovering the actual variations of $\bar{P_{\rm T}}^h$ with $e$.
However, the two layers model partly recovers the stagnation of $\bar{P_{\rm T}}^h$ at small $e$ whereas the exponential model qualitatively reproduces the kind of rise of $\bar{P_{\rm T}}^h$ towards its asymptote.
It thus appears that a mix of both models could be relevant.

We then conserve an homogeneous layer at particle density $\phi_2$ close to the sample plate and extend it, beyond a distance $\delta$, by a smooth exponential relaxation towards the limit density $\phi_3$.
The local volume fraction then reads :
\begin{eqnarray}
\nonumber
0 \leq y \leq \delta &:& \phi(y)=\phi_2, \\
\nonumber
\delta  \leq y  \leq e/2 &:& \phi(y)= \phi_3 + (\phi_2- \phi_3) \exp(-\frac{y-\delta}{\xi})
\end{eqnarray}
with $\delta$ the depth of the homogeneous ordered layer at the plate and $\xi$ the relaxation length of the particle density beyond.

The above analyses provide the following variation with $e$ of the inhomogeneity factor $I(\delta, \xi, e)$ :
\begin{eqnarray}
\nonumber
0 \leq e \leq 2 \delta &:& I(\delta,\xi,e)=\alpha, \\
\nonumber
2\delta  \leq e &:& I(\delta,\xi,e) \textrm{ as given by relation (\ref{I,exp})}
\end{eqnarray}
with the above values of $\beta$ and $\gamma$ but $\epsilon=(e/2 - \delta)/\xi$.
Note that $I$ is actually continuous since $I=\alpha$ for $\epsilon=0$ in relation (19).

The best fit of $\bar{P_{\rm T}}^h (e) = I(\xi,e)^{-1} \bar{P_{\rm T}}^i$ with fitting parameters $\delta$, $\xi$ and $\bar{P_{\rm T}}^i$ yields $\delta=24.4 \mu$m, $\xi = 9.9 \mu$m and $\bar{P_{\rm T}}^i = 2793$ Pa.
It recovers both the stagnation at small $e$ and the rise towards the asymptote.
The corresponding profiles of $\phi(y)$ and of $\bar{P_{\rm T}}^h (e)$ are reported on figures \ref{G,PT,e,Lin,Alle,fits} (a) and (b).

\subsection{Renormalization of the thermomolecular pressure}
\label{RenormalizationThermomolecularPressure}
The combined model thus satisfactorily recovers the data points $\bar{P_{\rm T}}^h (e)$ with a best fitting value of the actual thermomolecular pressure $\bar{P_{\rm T}}^i$ of $2793$ Pa.
It is however based on the guess that this thermomolecular pressure $\bar{P_{\rm T}}^i$ is independent of the sample depth $e$.
To evaluate its relevance and check the consistency of our approach, we determine \emph{a posteriori}, at each sample depth $e$, the value of $\bar{P_{\rm T}}^i (e)$ deduced from the data of $\bar{P_{\rm T}}^h (e)$, using relation (\ref{PT,i,h}) and the factor $I(\delta,\xi,e)$ (\ref{I}) with the best fitting parameters $\delta=24.4 \mu$m and $\xi = 9.9 \mu$m.

Figure \ref{G,PT,PTe} shows that the renormalized thermomolecular pressure $\bar{P_{\rm T}}^i (e)$ deduced this way is actually fairly constant. This contrasts with the large increase with $e$ of the value $\bar{P_{\rm T}}^h$ obtained when assuming a homogeneous layer.
This constancy solves the paradox of a thermomolecular pressure dependent on the sample depth by linking it to an incorrect assumption of particle layer homogeneity.
In addition, it provides a strong support to the relevance of the permeability mechanism to handle the effect of layer inhomogeneity.

\section{Discussion}
\label{Discussion}

Varying the sample depth $e$ in our experiment led us to evidence the implications of the inhomogeneity in particle density $\phi$  on the thickness $h$ of compacted particle layers.

The origin of this inhomogeneity lies in the ordering imposed by flat boundaries as also observed in granular materials.
Interestingly, although it is restricted to the plate vicinity, this inhomogeneity has proven to give rise here to macroscopic implication in the layer thickness.
This shows the global sensitivity of freezing suspensions to inhomogeneity in particle density and to confinement.

The layer thickness $h$, the Darcy velocity $U$ and the sample depth $e$ have been linked by a mechanical model of trapping involving the layer inhomogeneity.
The profile of $\phi$ that fits data the best at various sample depths consists, close to the plates, in several hexagonal close packed layers of maximum close packing fraction $\phi_2$ followed, further to the mid-depth, by an exponential relaxation towards the random close packing fraction $\phi_3$.

Although these conclusions were obtained at a definite particle diameter, they are expected to extend to other particle sizes insofar as no other phenomena such as electrostatic interactions interfere.
Accordingly, the quantitative determination of both the  depth $\delta$ of the sub-layer of density $\phi_2$ attached to the plate and the relaxation length $\xi$ towards $\phi_3$ beyond should better be considered in terms of particle diameter $d$.
This yields $\delta\approx 8 d$ and $\xi\approx 3 d$.
Both provide  a net influence length $\Lambda =\delta+\xi \approx 11 d$ of a smooth boundary on the particle density $\phi$.
In comparison the two layers model and the exponential model both led a shorter influence length of $\Lambda \approx 6d$.
Their poor fitting of data as compared to the combined model shows the relevance of an influence length as large as $11 d$ here.
This agrees with the influence length of flat plates in granular flows which has been found to extend over at least $5d$ \cite{Mandal2017}.
On the other hand, the small value of the relaxation length $\xi\approx 3 d$ as compared to the sub-layer length $\delta\approx 8 d$ supports a rather sharp nature of the transition from $\phi_2$ to $\phi_3$.
Direct analysis of the particle packing inside the suspension would be valuable to confirm this point.

On a general viewpoint, the transition from a random packing to an hexagonal packing when approaching the plate boundary corresponds to the formation of 2D colloidal crystals.
While crystallization can spontaneously occur in colloidal suspensions \cite{Pusey1989}, it has been found to be induced by shaking in granular materials \cite{Pouliquen1997}.
In suspensions, this ordering may be encountered during evaporation or drying \cite{Denkov1996} or by hydrodynamic shear \cite{Mueller2010,Ackerson1988,Wu2009}.
It is then mediated by capillary forces or by shear stresses respectively. 
Here, particles are compacted and pressed on the sample plate by the pressure exerted by the particle matrix.
This pressure seems necessary to induce crystallization since, on spheres packed by sedimentation in a fluidized bed, the particle density at the boundaries was found to be even smaller than in the bulk \cite{Jerkins2008}.
Here, this pressure increases from zero at the frontier between the compacted layer and the liquid phase to the pressure $\bar{P_{\rm p}}$ at the solidification front.
As it is weak in a part of the particle layer, it may thus not sufficiently compress particles on the plates to reach the density $\phi_2$ there.
We nevertheless assumed that this concerns a relatively negligible part of the layer.

Both the nature of random close packing and of the crystallization of sphere packing still raise open questions regarding the geometric frustration that prevents the former to cross the density limit $\phi_3$ and the particle rearrangements that enable crystal occurrence \cite{Francois2013}.
In particular, in random close packing, Bernal emphasized the relevance of local tetrahedral configurations yielding dense polytetrahedral  aggregates \cite{Bernal1964}.
These aggregates have indeed been related to geometric frustration \cite{Francois2013} and to crystallization by  disappearance of polyhedral clusters \cite{Anikeenko2007,Anikeenko2008}.
Although a noticeable support has been provided on these views and on their implications on geometric frustration and on crystallization, these issues still remain in debate.
The present suspension indicates that, in presence of Darcy flows, the transition from random close packing to hexagonal crystal is somewhat sharp since it is completed over about three particles.
Whether this is compatible with actual views on crystallization will be a stimulating issue to address.
	
The findings of this study refer more generally to the implication of smooth boundaries on the organization of suspensions.
It then appears that the smoothness of boundaries may promote crystallization of sufficiently compacted suspensions over a distance that depends on the particle diameter.
We note that this mechanism might apply to polydisperse suspensions involving particles of largely different sizes insofar as large particles play the role of smooth boundaries for the small ones.
Then, noticeable implications on the crystallization of small particles may be expected if the mean distance between large particles is close to or below the influence length of smooth boundaries on small particles.
In solidifying bimodal suspensions, this should affect the density of small particles with noticeable mechanical implications on the permeability coefficient of the compacted particle layer, its thickness and its morphological stability.
In the sedimentation of suspensions, a similar effect might yield a boundary-induced crystallization of small particles either near sample boundaries or in between large particles.

On the other hand, the crystallization process of colloidal suspensions has displayed a large sensitivity to polydispersity \cite{Schope2006}.
In particular, small changes of the particle size distribution have been found to induce both delay and enhancement of crystal nucleation as a result of particle fractionation \cite{Martin2003}.
Here, similar mechanisms might play a relevant role in the boundary-induced crystallization close to smooth boundaries.
Investigating these mechanisms in the context of solidifying suspensions crossed by Darcy flows should thus provide relevant insights into the spatial and temporal formation of hexagonal close packing close to smooth boundaries. 
This could then help understanding the main features pointed out in this study : the influence length of soft boundaries and the nature of the transition from hexagonal close packing close to them to random close packing farther into the bulk.

\section{Conclusion}
\label{Conclusion}

Freezing monodisperse suspensions in directional experiments in thin samples has evidenced that the thickness of the compacted layer formed ahead of the solidification front increases with the sample depth at other parameters fixed.
This effect has been related to the inhomogeneity of particle density brought about by the long range correlation order imposed by the sample plates.
This leads the particle density to evolve from that of hexagonal packing at the sample plate to that of random close packing far from it.

A mechanical model of trapping has provided a link between the layer thickness, the particle inhomogeneity and the sample depth.
It is based on the sensitivity of the layer permeability to the particle density.
Following its validation at the extreme sample depths where layers can be approximated as homogeneous, this "permeability mechanism" has been used to deduce the particle profile that best fits the layer thickness evolution.
This profile enabled a constant thermomolecular pressure to be recovered and revealed a somewhat large influence length of smooth boundaries of about $11$ particle diameters.

Beyond the deepening of the physical analysis of freezing suspensions and the clarification of the implications of boundaries on them, this study may be useful to better model bimodal suspensions when large particles act as smooth boundaries for small particles, not only in solidification but also in sedimentation.
More generally, it shows that pattern formation in freezing suspensions also refers to particle organization from crystallized to random packed.

\section*{Acknowledgments}
The research leading to these results has been supported by the European Research Council under the European Union's Seventh Framework Program (FP7/2007-2013)/ERC Grant Agreement No. 278004.
    
\bibliography{ReferenceBoundaryInducedInhomogeneity}

\begin{thebibliography}{68}
\expandafter\ifx\csname natexlab\endcsname\relax\def\natexlab#1{#1}\fi
\expandafter\ifx\csname bibnamefont\endcsname\relax
  \def\bibnamefont#1{#1}\fi
\expandafter\ifx\csname bibfnamefont\endcsname\relax
  \def\bibfnamefont#1{#1}\fi
\expandafter\ifx\csname citenamefont\endcsname\relax
  \def\citenamefont#1{#1}\fi
\expandafter\ifx\csname url\endcsname\relax
  \def\url#1{\texttt{#1}}\fi
\expandafter\ifx\csname urlprefix\endcsname\relax\def\urlprefix{URL }\fi
\providecommand{\bibinfo}[2]{#2}
\providecommand{\eprint}[2][]{\url{#2}}

\bibitem[{\citenamefont{Zhu et~al.}(2000)\citenamefont{Zhu, Vilches, Dash,
  Sing, and Wettlaufer}}]{Zhu2000}
\bibinfo{author}{\bibfnamefont{D.~M.} \bibnamefont{Zhu}},
  \bibinfo{author}{\bibfnamefont{O.~E.} \bibnamefont{Vilches}},
  \bibinfo{author}{\bibfnamefont{J.~G.} \bibnamefont{Dash}},
  \bibinfo{author}{\bibfnamefont{B.}~\bibnamefont{Sing}}, \bibnamefont{and}
  \bibinfo{author}{\bibfnamefont{J.~S.} \bibnamefont{Wettlaufer}},
  \bibinfo{journal}{Physical Review Letters} \textbf{\bibinfo{volume}{85}},
  \bibinfo{pages}{4908} (\bibinfo{year}{2000}).

\bibitem[{\citenamefont{Rempel et~al.}(2004)\citenamefont{Rempel, Wettlaufer,
  and Worster}}]{Rempel2004}
\bibinfo{author}{\bibfnamefont{A.~W.} \bibnamefont{Rempel}},
  \bibinfo{author}{\bibfnamefont{J.~S.} \bibnamefont{Wettlaufer}},
  \bibnamefont{and} \bibinfo{author}{\bibfnamefont{M.~G.}
  \bibnamefont{Worster}}, \bibinfo{journal}{Journal of Fluid Mechanics}
  \textbf{\bibinfo{volume}{498}}, \bibinfo{pages}{227} (\bibinfo{year}{2004}).

\bibitem[{\citenamefont{Peppin and Style}(2013)}]{Peppin2013}
\bibinfo{author}{\bibfnamefont{S.~S.~L.} \bibnamefont{Peppin}}
  \bibnamefont{and} \bibinfo{author}{\bibfnamefont{R.~W.} \bibnamefont{Style}},
  \bibinfo{journal}{Vadose Zone Journal} \textbf{\bibinfo{volume}{12}}
  (\bibinfo{year}{2013}).

\bibitem[{\citenamefont{Mutou et~al.}(1998)\citenamefont{Mutou, Watanabe,
  Ishizaki, and Mizoguchi}}]{Mutou1998}
\bibinfo{author}{\bibfnamefont{Y.}~\bibnamefont{Mutou}},
  \bibinfo{author}{\bibfnamefont{K.}~\bibnamefont{Watanabe}},
  \bibinfo{author}{\bibfnamefont{T.}~\bibnamefont{Ishizaki}}, \bibnamefont{and}
  \bibinfo{author}{\bibfnamefont{M.}~\bibnamefont{Mizoguchi}},
  \bibinfo{journal}{Proc. 7th Intl Conf. on Permafrost, Yellowknife, Canada,
  Centre d'Etudes Nordique, Universit\'e Laval, Canada} pp.
  \bibinfo{pages}{783--787} (\bibinfo{year}{1998}).

\bibitem[{\citenamefont{Saruya et~al.}(2013)\citenamefont{Saruya, Kurita, and
  Rempel}}]{Saruya2013}
\bibinfo{author}{\bibfnamefont{T.}~\bibnamefont{Saruya}},
  \bibinfo{author}{\bibfnamefont{K.}~\bibnamefont{Kurita}}, \bibnamefont{and}
  \bibinfo{author}{\bibfnamefont{A.~W.} \bibnamefont{Rempel}},
  \bibinfo{journal}{Physical Review E} \textbf{\bibinfo{volume}{87}},
  \bibinfo{pages}{032404} (\bibinfo{year}{2013}).

\bibitem[{\citenamefont{Anderson and {Grae Worster}}(2014)}]{Anderson2014}
\bibinfo{author}{\bibfnamefont{A.~M.} \bibnamefont{Anderson}} \bibnamefont{and}
  \bibinfo{author}{\bibfnamefont{M.}~\bibnamefont{{Grae Worster}}},
  \bibinfo{journal}{Journal of Fluid Mechanics} \textbf{\bibinfo{volume}{758}},
  \bibinfo{pages}{786} (\bibinfo{year}{2014}).

\bibitem[{\citenamefont{Ping et~al.}(2008)\citenamefont{Ping, Michaelson,
  Kimble, Romanovsky, Shur, Swanson, and Walker}}]{Ping2008}
\bibinfo{author}{\bibfnamefont{C.~L.} \bibnamefont{Ping}},
  \bibinfo{author}{\bibfnamefont{G.~J.} \bibnamefont{Michaelson}},
  \bibinfo{author}{\bibfnamefont{J.~M.} \bibnamefont{Kimble}},
  \bibinfo{author}{\bibfnamefont{V.~E.} \bibnamefont{Romanovsky}},
  \bibinfo{author}{\bibfnamefont{Y.~L.} \bibnamefont{Shur}},
  \bibinfo{author}{\bibfnamefont{D.~K.} \bibnamefont{Swanson}},
  \bibnamefont{and} \bibinfo{author}{\bibfnamefont{D.~A.}
  \bibnamefont{Walker}}, \bibinfo{journal}{Journal of Geophysical Research}
  \textbf{\bibinfo{volume}{113}}, \bibinfo{pages}{G03S12}
  (\bibinfo{year}{2008}).

\bibitem[{\citenamefont{Rahman and Velez-Ruiz}(2007)}]{Velez-Ruiz2007}
\bibinfo{author}{\bibfnamefont{M.~S.} \bibnamefont{Rahman}} \bibnamefont{and}
  \bibinfo{author}{\bibfnamefont{J.~F.} \bibnamefont{Velez-Ruiz}},
  \bibinfo{journal}{CDC Press: Boca Raton, FL} pp. \bibinfo{pages}{635--665}
  (\bibinfo{year}{2007}).

\bibitem[{\citenamefont{Bronstein et~al.}(1981)\citenamefont{Bronstein, Itkin,
  and Ishkov}}]{Bronstein1981}
\bibinfo{author}{\bibfnamefont{V.~L.} \bibnamefont{Bronstein}},
  \bibinfo{author}{\bibfnamefont{Y.~A.} \bibnamefont{Itkin}}, \bibnamefont{and}
  \bibinfo{author}{\bibfnamefont{G.~S.} \bibnamefont{Ishkov}},
  \bibinfo{journal}{Journal of Crystal Growth} \textbf{\bibinfo{volume}{52}},
  \bibinfo{pages}{345} (\bibinfo{year}{1981}).

\bibitem[{\citenamefont{K{\"{o}}rber}(1988)}]{Korber1988}
\bibinfo{author}{\bibfnamefont{C.}~\bibnamefont{K{\"{o}}rber}},
  \bibinfo{journal}{Quarterly reviews of biophysics}
  \textbf{\bibinfo{volume}{21}}, \bibinfo{pages}{229} (\bibinfo{year}{1988}).

\bibitem[{\citenamefont{Muldrew et~al.}(2000)\citenamefont{Muldrew, Novak,
  Yang, Zernicke, Schachar, and McGann}}]{Muldrew2000}
\bibinfo{author}{\bibfnamefont{K.}~\bibnamefont{Muldrew}},
  \bibinfo{author}{\bibfnamefont{K.}~\bibnamefont{Novak}},
  \bibinfo{author}{\bibfnamefont{H.}~\bibnamefont{Yang}},
  \bibinfo{author}{\bibfnamefont{R.}~\bibnamefont{Zernicke}},
  \bibinfo{author}{\bibfnamefont{N.~S.} \bibnamefont{Schachar}},
  \bibnamefont{and} \bibinfo{author}{\bibfnamefont{L.~E.}
  \bibnamefont{McGann}}, \bibinfo{journal}{Cryobiology}
  \textbf{\bibinfo{volume}{40}}, \bibinfo{pages}{102 } (\bibinfo{year}{2000}).

\bibitem[{\citenamefont{Stefanescu et~al.}(1988)\citenamefont{Stefanescu,
  Dhindaw, Kacar, and Moitra}}]{Stefanescu1988}
\bibinfo{author}{\bibfnamefont{D.~M.} \bibnamefont{Stefanescu}},
  \bibinfo{author}{\bibfnamefont{B.~K.} \bibnamefont{Dhindaw}},
  \bibinfo{author}{\bibfnamefont{S.~A.} \bibnamefont{Kacar}}, \bibnamefont{and}
  \bibinfo{author}{\bibfnamefont{A.}~\bibnamefont{Moitra}},
  \bibinfo{journal}{Metallurgical Transactions A}
  \textbf{\bibinfo{volume}{19}}, \bibinfo{pages}{2847} (\bibinfo{year}{1988}).

\bibitem[{\citenamefont{Deville et~al.}(2007)\citenamefont{Deville, Saiz, and
  Tomsia}}]{Deville2007a}
\bibinfo{author}{\bibfnamefont{S.}~\bibnamefont{Deville}},
  \bibinfo{author}{\bibfnamefont{E.}~\bibnamefont{Saiz}}, \bibnamefont{and}
  \bibinfo{author}{\bibfnamefont{A.~P.} \bibnamefont{Tomsia}},
  \bibinfo{journal}{Acta Materialia} \textbf{\bibinfo{volume}{55}},
  \bibinfo{pages}{1965} (\bibinfo{year}{2007}).

\bibitem[{\citenamefont{Uhlmann et~al.}(1964)\citenamefont{Uhlmann, Chalmers,
  and Jackson}}]{Uhlmann1964}
\bibinfo{author}{\bibfnamefont{D.~R.} \bibnamefont{Uhlmann}},
  \bibinfo{author}{\bibfnamefont{B.}~\bibnamefont{Chalmers}}, \bibnamefont{and}
  \bibinfo{author}{\bibfnamefont{K.~A.} \bibnamefont{Jackson}},
  \bibinfo{journal}{Journal of Applied Physics} \textbf{\bibinfo{volume}{35}},
  \bibinfo{pages}{2986} (\bibinfo{year}{1964}).

\bibitem[{\citenamefont{Ciss{\'{e}}}(1971)}]{Cisse1971}
\bibinfo{author}{\bibfnamefont{J.}~\bibnamefont{Ciss{\'{e}}}},
  \bibinfo{journal}{Journal of Crystal Growth} \textbf{\bibinfo{volume}{10}},
  \bibinfo{pages}{67} (\bibinfo{year}{1971}).

\bibitem[{\citenamefont{Zubko et~al.}(1973)\citenamefont{Zubko, Lobanov, and
  Nikonova}}]{Zubko1973}
\bibinfo{author}{\bibfnamefont{A.~M.} \bibnamefont{Zubko}},
  \bibinfo{author}{\bibfnamefont{V.~G.} \bibnamefont{Lobanov}},
  \bibnamefont{and} \bibinfo{author}{\bibfnamefont{V.~V.}
  \bibnamefont{Nikonova}}, \bibinfo{journal}{Soviet Physics and
  Crystallography} \textbf{\bibinfo{volume}{18}}, \bibinfo{pages}{239}
  (\bibinfo{year}{1973}).

\bibitem[{\citenamefont{Chernov et~al.}(1976)\citenamefont{Chernov, Temkin, and
  Mel'nikova}}]{Chernov1976}
\bibinfo{author}{\bibfnamefont{A.~A.} \bibnamefont{Chernov}},
  \bibinfo{author}{\bibfnamefont{D.~E.} \bibnamefont{Temkin}},
  \bibnamefont{and} \bibinfo{author}{\bibfnamefont{A.~M.}
  \bibnamefont{Mel'nikova}}, \bibinfo{journal}{Soviet Physics and
  Crystallography} \textbf{\bibinfo{volume}{21}}, \bibinfo{pages}{369}
  (\bibinfo{year}{1976}).

\bibitem[{\citenamefont{K{\"{o}}rber et~al.}(1985)\citenamefont{K{\"{o}}rber,
  Rau, Cosman, and Cravalho}}]{Korber1985}
\bibinfo{author}{\bibfnamefont{C.}~\bibnamefont{K{\"{o}}rber}},
  \bibinfo{author}{\bibfnamefont{G.}~\bibnamefont{Rau}},
  \bibinfo{author}{\bibfnamefont{M.}~\bibnamefont{Cosman}}, \bibnamefont{and}
  \bibinfo{author}{\bibfnamefont{E.}~\bibnamefont{Cravalho}},
  \bibinfo{journal}{Journal of crystal growth} \textbf{\bibinfo{volume}{72}},
  \bibinfo{pages}{649} (\bibinfo{year}{1985}).

\bibitem[{\citenamefont{Lipp et~al.}(1990)\citenamefont{Lipp, K{\"{o}}rber, and
  Rau}}]{Lipp1990}
\bibinfo{author}{\bibfnamefont{G.}~\bibnamefont{Lipp}},
  \bibinfo{author}{\bibfnamefont{C.}~\bibnamefont{K{\"{o}}rber}},
  \bibnamefont{and} \bibinfo{author}{\bibfnamefont{G.}~\bibnamefont{Rau}},
  \bibinfo{journal}{Journal of crystal growth} \textbf{\bibinfo{volume}{99}},
  \bibinfo{pages}{206} (\bibinfo{year}{1990}).

\bibitem[{\citenamefont{Lipp and K{\"{o}}rber}(1993)}]{Lipp1993}
\bibinfo{author}{\bibfnamefont{G.}~\bibnamefont{Lipp}} \bibnamefont{and}
  \bibinfo{author}{\bibfnamefont{C.}~\bibnamefont{K{\"{o}}rber}},
  \bibinfo{journal}{Journal of crystal growth} \textbf{\bibinfo{volume}{130}},
  \bibinfo{pages}{475} (\bibinfo{year}{1993}).

\bibitem[{\citenamefont{Rempel and Worster}(1999)}]{Rempel1999}
\bibinfo{author}{\bibfnamefont{A.~W.} \bibnamefont{Rempel}} \bibnamefont{and}
  \bibinfo{author}{\bibfnamefont{M.~G.} \bibnamefont{Worster}},
  \bibinfo{journal}{Journal of Crystal Growth} \textbf{\bibinfo{volume}{205}},
  \bibinfo{pages}{427} (\bibinfo{year}{1999}).

\bibitem[{\citenamefont{Rempel and Worster}(2001)}]{Rempel2001}
\bibinfo{author}{\bibfnamefont{A.~W.} \bibnamefont{Rempel}} \bibnamefont{and}
  \bibinfo{author}{\bibfnamefont{M.~G.} \bibnamefont{Worster}},
  \bibinfo{journal}{Journal of Crystal Growth} \textbf{\bibinfo{volume}{223}},
  \bibinfo{pages}{420} (\bibinfo{year}{2001}).

\bibitem[{\citenamefont{Park et~al.}(2006)\citenamefont{Park, Golovin, and
  Davis}}]{Park2006}
\bibinfo{author}{\bibfnamefont{M.~S.} \bibnamefont{Park}},
  \bibinfo{author}{\bibfnamefont{A.~A.} \bibnamefont{Golovin}},
  \bibnamefont{and} \bibinfo{author}{\bibfnamefont{S.~H.} \bibnamefont{Davis}},
  \bibinfo{journal}{Journal of Fluid Mechanics} \textbf{\bibinfo{volume}{560}},
  \bibinfo{pages}{415} (\bibinfo{year}{2006}).

\bibitem[{\citenamefont{Dash et~al.}(2006)\citenamefont{Dash, Rempel, and
  Wettlaufer}}]{Dash2006}
\bibinfo{author}{\bibfnamefont{J.}~\bibnamefont{Dash}},
  \bibinfo{author}{\bibfnamefont{A.~W.} \bibnamefont{Rempel}},
  \bibnamefont{and} \bibinfo{author}{\bibfnamefont{J.~S.}
  \bibnamefont{Wettlaufer}}, \bibinfo{journal}{Reviews of Modern Physics}
  \textbf{\bibinfo{volume}{78}}, \bibinfo{pages}{695} (\bibinfo{year}{2006}).

\bibitem[{\citenamefont{Peppin et~al.}(2006)\citenamefont{Peppin, Elliott, and
  Worster}}]{Peppin2006}
\bibinfo{author}{\bibfnamefont{S.~S.~L.} \bibnamefont{Peppin}},
  \bibinfo{author}{\bibfnamefont{J.~A.~W.} \bibnamefont{Elliott}},
  \bibnamefont{and} \bibinfo{author}{\bibfnamefont{M.~G.}
  \bibnamefont{Worster}}, \bibinfo{journal}{Journal of Fluid Mechanics}
  \textbf{\bibinfo{volume}{554}}, \bibinfo{pages}{147} (\bibinfo{year}{2006}).

\bibitem[{\citenamefont{Peppin et~al.}(2008)\citenamefont{Peppin, Wettlaufer,
  and Worster}}]{Peppin2008}
\bibinfo{author}{\bibfnamefont{S.~S.~L.} \bibnamefont{Peppin}},
  \bibinfo{author}{\bibfnamefont{J.~S.} \bibnamefont{Wettlaufer}},
  \bibnamefont{and} \bibinfo{author}{\bibfnamefont{M.~G.}
  \bibnamefont{Worster}}, \bibinfo{journal}{Phys. Rev. lett.}
  \textbf{\bibinfo{volume}{100}}, \bibinfo{pages}{238301}
  (\bibinfo{year}{2008}).

\bibitem[{\citenamefont{Anderson and Worster}(2012)}]{Anderson2012}
\bibinfo{author}{\bibfnamefont{A.~M.} \bibnamefont{Anderson}} \bibnamefont{and}
  \bibinfo{author}{\bibfnamefont{M.~G.} \bibnamefont{Worster}},
  \bibinfo{journal}{Langmuir} \textbf{\bibinfo{volume}{28}},
  \bibinfo{pages}{16512} (\bibinfo{year}{2012}).

\bibitem[{\citenamefont{Saint-Michel et~al.}(2017)\citenamefont{Saint-Michel,
  Georgelin, Deville, and Pocheau}}]{Saint-Michel2017}
\bibinfo{author}{\bibfnamefont{B.}~\bibnamefont{Saint-Michel}},
  \bibinfo{author}{\bibfnamefont{M.}~\bibnamefont{Georgelin}},
  \bibinfo{author}{\bibfnamefont{S.}~\bibnamefont{Deville}}, \bibnamefont{and}
  \bibinfo{author}{\bibfnamefont{A.}~\bibnamefont{Pocheau}},
  \bibinfo{journal}{Langmuir} \textbf{\bibinfo{volume}{33}},
  \bibinfo{pages}{5617} (\bibinfo{year}{2017}).

\bibitem[{\citenamefont{Bridgman}(1925)}]{Bridgman1925}
\bibinfo{author}{\bibfnamefont{P.~W.} \bibnamefont{Bridgman}},
  \bibinfo{journal}{Proceedings of the American Academy of Arts and Sciences}
  \textbf{\bibinfo{volume}{60}}, \bibinfo{pages}{305} (\bibinfo{year}{1925}).

\bibitem[{\citenamefont{Stockbarger}(1936)}]{Stockbarger1936}
\bibinfo{author}{\bibfnamefont{D.}~\bibnamefont{Stockbarger}},
  \bibinfo{journal}{Review of Scientific Instruments}
  \textbf{\bibinfo{volume}{7}}, \bibinfo{pages}{133} (\bibinfo{year}{1936}).

\bibitem[{\citenamefont{Hunt et~al.}(1966)\citenamefont{Hunt, Jackson, and
  Brown}}]{Hunt1966}
\bibinfo{author}{\bibfnamefont{D.}~\bibnamefont{Hunt}},
  \bibinfo{author}{\bibfnamefont{K.}~\bibnamefont{Jackson}}, \bibnamefont{and}
  \bibinfo{author}{\bibfnamefont{H.}~\bibnamefont{Brown}},
  \bibinfo{journal}{Rev. Sci. Instrum.} \textbf{\bibinfo{volume}{37}},
  \bibinfo{pages}{805} (\bibinfo{year}{1966}).

\bibitem[{\citenamefont{Georgelin and Pocheau}(1998)}]{Georgelin1998}
\bibinfo{author}{\bibfnamefont{M.}~\bibnamefont{Georgelin}} \bibnamefont{and}
  \bibinfo{author}{\bibfnamefont{A.}~\bibnamefont{Pocheau}},
  \bibinfo{journal}{Phys. Rev. E} \textbf{\bibinfo{volume}{57}},
  \bibinfo{pages}{3189} (\bibinfo{year}{1998}).

\bibitem[{\citenamefont{Pocheau and Georgelin}(2006)}]{Pocheau2006}
\bibinfo{author}{\bibfnamefont{A.}~\bibnamefont{Pocheau}} \bibnamefont{and}
  \bibinfo{author}{\bibfnamefont{M.}~\bibnamefont{Georgelin}},
  \bibinfo{journal}{Phys.Rev.E} \textbf{\bibinfo{volume}{73}},
  \bibinfo{pages}{011604} (\bibinfo{year}{2006}).

\bibitem[{\citenamefont{Deschamps et~al.}(2006)\citenamefont{Deschamps,
  Georgelin, and Pocheau}}]{Deschamps2006}
\bibinfo{author}{\bibfnamefont{J.}~\bibnamefont{Deschamps}},
  \bibinfo{author}{\bibfnamefont{M.}~\bibnamefont{Georgelin}},
  \bibnamefont{and} \bibinfo{author}{\bibfnamefont{A.}~\bibnamefont{Pocheau}},
  \bibinfo{journal}{Euro. Phys. Lett.} \textbf{\bibinfo{volume}{76}},
  \bibinfo{pages}{291} (\bibinfo{year}{2006}).

\bibitem[{\citenamefont{Beckmann et~al.}(1990)\citenamefont{Beckmann,
  K{\"{o}}rber, Rau, Hubel, and Cravalho}}]{Beckmann1990}
\bibinfo{author}{\bibfnamefont{J.}~\bibnamefont{Beckmann}},
  \bibinfo{author}{\bibfnamefont{C.}~\bibnamefont{K{\"{o}}rber}},
  \bibinfo{author}{\bibfnamefont{G.}~\bibnamefont{Rau}},
  \bibinfo{author}{\bibfnamefont{A.}~\bibnamefont{Hubel}}, \bibnamefont{and}
  \bibinfo{author}{\bibfnamefont{E.}~\bibnamefont{Cravalho}},
  \bibinfo{journal}{Cryobiology} \textbf{\bibinfo{volume}{27}},
  \bibinfo{pages}{279} (\bibinfo{year}{1990}).

\bibitem[{\citenamefont{Pocheau et~al.}(2009)\citenamefont{Pocheau, Bodea, and
  Georgelin}}]{Pocheau2009}
\bibinfo{author}{\bibfnamefont{A.}~\bibnamefont{Pocheau}},
  \bibinfo{author}{\bibfnamefont{S.}~\bibnamefont{Bodea}}, \bibnamefont{and}
  \bibinfo{author}{\bibfnamefont{M.}~\bibnamefont{Georgelin}},
  \bibinfo{journal}{Phys. Rev. E} \textbf{\bibinfo{volume}{80}},
  \bibinfo{pages}{031601} (\bibinfo{year}{2009}).

\bibitem[{\citenamefont{Song et~al.}(2008)\citenamefont{Song, Wang, and
  Makse}}]{Song2008}
\bibinfo{author}{\bibfnamefont{C.}~\bibnamefont{Song}},
  \bibinfo{author}{\bibfnamefont{P.}~\bibnamefont{Wang}}, \bibnamefont{and}
  \bibinfo{author}{\bibfnamefont{H.~A.} \bibnamefont{Makse}},
  \bibinfo{journal}{Nature} \textbf{\bibinfo{volume}{453}},
  \bibinfo{pages}{629} (\bibinfo{year}{2008}).

\bibitem[{\citenamefont{Janssen}(1895)}]{Janssen1895}
\bibinfo{author}{\bibfnamefont{H.}~\bibnamefont{Janssen}},
  \bibinfo{journal}{Zeitschr. d. Vereines Deutscher Ingenieure}
  \textbf{\bibinfo{volume}{39}}, \bibinfo{pages}{1045} (\bibinfo{year}{1895}).

\bibitem[{\citenamefont{Sperl}(2006)}]{JanssenSperl1895}
\bibinfo{author}{\bibfnamefont{M.}~\bibnamefont{Sperl}},
  \bibinfo{journal}{Granular Matter} \textbf{\bibinfo{volume}{8}},
  \bibinfo{pages}{59} (\bibinfo{year}{2006}).

\bibitem[{\citenamefont{Andreotti et~al.}(2013)\citenamefont{Andreotti,
  Forterre, and Pouliquen}}]{AndreottiForterrePouliquen2013}
\bibinfo{author}{\bibfnamefont{B.}~\bibnamefont{Andreotti}},
  \bibinfo{author}{\bibfnamefont{Y.}~\bibnamefont{Forterre}}, \bibnamefont{and}
  \bibinfo{author}{\bibfnamefont{O.}~\bibnamefont{Pouliquen}},
  \emph{\bibinfo{title}{Granular Media: Between Fluid and Solid}}
  (\bibinfo{publisher}{Cambridge University Press}, \bibinfo{year}{2013}).

\bibitem[{\citenamefont{Saint-Michel et~al.}(2018)\citenamefont{Saint-Michel,
  Georgelin, Deville, and Pocheau}}]{Saint-Michel2018}
\bibinfo{author}{\bibfnamefont{B.}~\bibnamefont{Saint-Michel}},
  \bibinfo{author}{\bibfnamefont{M.}~\bibnamefont{Georgelin}},
  \bibinfo{author}{\bibfnamefont{S.}~\bibnamefont{Deville}}, \bibnamefont{and}
  \bibinfo{author}{\bibfnamefont{A.}~\bibnamefont{Pocheau}},
  \bibinfo{journal}{Soft Matter} \textbf{\bibinfo{volume}{14}},
  \bibinfo{pages}{9498} (\bibinfo{year}{2018}).

\bibitem[{\citenamefont{Israelachvili}(1991)}]{Israelachvili1992Book}
\bibinfo{author}{\bibfnamefont{J.~N.} \bibnamefont{Israelachvili}},
  \emph{\bibinfo{title}{Intermolecular and surface forces / Jacob N.
  Israelachvili}} (\bibinfo{publisher}{Academic Press London ; San Diego},
  \bibinfo{year}{1991}), \bibinfo{edition}{2nd} ed., ISBN
  \bibinfo{isbn}{0123751810}.

\bibitem[{\citenamefont{Wilen et~al.}(1995)\citenamefont{Wilen, Wettlaufer,
  Elbaum, and Schick}}]{Wilen1995}
\bibinfo{author}{\bibfnamefont{L.~A.} \bibnamefont{Wilen}},
  \bibinfo{author}{\bibfnamefont{J.~S.} \bibnamefont{Wettlaufer}},
  \bibinfo{author}{\bibfnamefont{M.}~\bibnamefont{Elbaum}}, \bibnamefont{and}
  \bibinfo{author}{\bibfnamefont{M.}~\bibnamefont{Schick}},
  \bibinfo{journal}{Phys. Rev. B} \textbf{\bibinfo{volume}{52}},
  \bibinfo{pages}{12426} (\bibinfo{year}{1995}).

\bibitem[{\citenamefont{Wettlaufer}(1999)}]{Wettlaufer1999}
\bibinfo{author}{\bibfnamefont{J.~S.} \bibnamefont{Wettlaufer}},
  \bibinfo{journal}{Physical Review Letters} \textbf{\bibinfo{volume}{82}},
  \bibinfo{pages}{2516} (\bibinfo{year}{1999}).

\bibitem[{\citenamefont{Wettlaufer and Worster}(2006)}]{Wettlaufer2006}
\bibinfo{author}{\bibfnamefont{J.~S.} \bibnamefont{Wettlaufer}}
  \bibnamefont{and} \bibinfo{author}{\bibfnamefont{M.~G.}
  \bibnamefont{Worster}}, \bibinfo{journal}{Annual Review of Fluid Mechanics}
  \textbf{\bibinfo{volume}{38}}, \bibinfo{pages}{427} (\bibinfo{year}{2006}).

\bibitem[{\citenamefont{de~Gennes}(1999)}]{deGennes1999}
\bibinfo{author}{\bibfnamefont{P.}~\bibnamefont{de~Gennes}},
  \bibinfo{journal}{Reviews of Modern Physics} \textbf{\bibinfo{volume}{71}},
  \bibinfo{pages}{S374} (\bibinfo{year}{1999}).

\bibitem[{\citenamefont{Hales}(2005)}]{Hales2005}
\bibinfo{author}{\bibfnamefont{T.~C.} \bibnamefont{Hales}},
  \bibinfo{journal}{Ann. of Math.} \textbf{\bibinfo{volume}{162}},
  \bibinfo{pages}{1065} (\bibinfo{year}{2005}).

\bibitem[{\citenamefont{Scott and Kilgour}(1969)}]{Scott1969}
\bibinfo{author}{\bibfnamefont{G.~D.} \bibnamefont{Scott}} \bibnamefont{and}
  \bibinfo{author}{\bibfnamefont{D.~M.} \bibnamefont{Kilgour}},
  \bibinfo{journal}{Journal of Physics D: Applied Physics}
  \textbf{\bibinfo{volume}{2}}, \bibinfo{pages}{863} (\bibinfo{year}{1969}).

\bibitem[{\citenamefont{Torquato et~al.}(2000)\citenamefont{Torquato, Truskett,
  and Debenedetti}}]{Torquado2000}
\bibinfo{author}{\bibfnamefont{S.}~\bibnamefont{Torquato}},
  \bibinfo{author}{\bibfnamefont{T.~M.} \bibnamefont{Truskett}},
  \bibnamefont{and} \bibinfo{author}{\bibfnamefont{P.~G.}
  \bibnamefont{Debenedetti}}, \bibinfo{journal}{Phys. Rev. Lett.}
  \textbf{\bibinfo{volume}{84}}, \bibinfo{pages}{2064} (\bibinfo{year}{2000}).

\bibitem[{\citenamefont{Zhang et~al.}(2006)\citenamefont{Zhang, Thompson, Reed,
  and Beeken}}]{Zhang2006}
\bibinfo{author}{\bibfnamefont{W.}~\bibnamefont{Zhang}},
  \bibinfo{author}{\bibfnamefont{K.}~\bibnamefont{Thompson}},
  \bibinfo{author}{\bibfnamefont{A.}~\bibnamefont{Reed}}, \bibnamefont{and}
  \bibinfo{author}{\bibfnamefont{L.}~\bibnamefont{Beeken}},
  \bibinfo{journal}{Chemical Engineering Science}
  \textbf{\bibinfo{volume}{61}}, \bibinfo{pages}{8060} (\bibinfo{year}{2006}),
  ISSN \bibinfo{issn}{0009-2509}.

\bibitem[{\citenamefont{Burtseva et~al.}(2015)\citenamefont{Burtseva,
  Valdez~Salas, Werner, and Petranovskii}}]{Burtseva2015}
\bibinfo{author}{\bibfnamefont{L.}~\bibnamefont{Burtseva}},
  \bibinfo{author}{\bibfnamefont{B.}~\bibnamefont{Valdez~Salas}},
  \bibinfo{author}{\bibfnamefont{F.}~\bibnamefont{Werner}}, \bibnamefont{and}
  \bibinfo{author}{\bibfnamefont{V.}~\bibnamefont{Petranovskii}},
  \bibinfo{journal}{Revista Mexicana de F\'{\i}sica}
  \textbf{\bibinfo{volume}{61}}, \bibinfo{pages}{20} (\bibinfo{year}{2015}).

\bibitem[{\citenamefont{Mandal and Khakhar}(2017)}]{Mandal2017}
\bibinfo{author}{\bibfnamefont{S.}~\bibnamefont{Mandal}} \bibnamefont{and}
  \bibinfo{author}{\bibfnamefont{D.~V.} \bibnamefont{Khakhar}},
  \bibinfo{journal}{Physical Review E} \textbf{\bibinfo{volume}{96}},
  \bibinfo{pages}{050901} (\bibinfo{year}{2017}).

\bibitem[{\citenamefont{Desmond and Weeks}(2009)}]{Desmond2009}
\bibinfo{author}{\bibfnamefont{K.~W.} \bibnamefont{Desmond}} \bibnamefont{and}
  \bibinfo{author}{\bibfnamefont{E.~R.} \bibnamefont{Weeks}},
  \bibinfo{journal}{Phys. Rev. E} \textbf{\bibinfo{volume}{80}},
  \bibinfo{pages}{051305} (\bibinfo{year}{2009}).

\bibitem[{\citenamefont{Mueth}(2003)}]{Mueth2003}
\bibinfo{author}{\bibfnamefont{D.~M.} \bibnamefont{Mueth}},
  \bibinfo{journal}{Phys. Rev. E} \textbf{\bibinfo{volume}{67}},
  \bibinfo{pages}{011304} (\bibinfo{year}{2003}).

\bibitem[{\citenamefont{Pouliquen et~al.}(1997)\citenamefont{Pouliquen,
  Nicolas, and Weidman}}]{Pouliquen1997}
\bibinfo{author}{\bibfnamefont{O.}~\bibnamefont{Pouliquen}},
  \bibinfo{author}{\bibfnamefont{M.}~\bibnamefont{Nicolas}}, \bibnamefont{and}
  \bibinfo{author}{\bibfnamefont{P.~D.} \bibnamefont{Weidman}},
  \bibinfo{journal}{Physical Review Letters} \textbf{\bibinfo{volume}{79}},
  \bibinfo{pages}{3640} (\bibinfo{year}{1997}).

\bibitem[{\citenamefont{Bertho et~al.}(2003)\citenamefont{Bertho,
  Giorgiutti-Dauphin\'e, and Hulin}}]{Bertho2003}
\bibinfo{author}{\bibfnamefont{Y.}~\bibnamefont{Bertho}},
  \bibinfo{author}{\bibfnamefont{F.}~\bibnamefont{Giorgiutti-Dauphin\'e}},
  \bibnamefont{and} \bibinfo{author}{\bibfnamefont{J.-P.} \bibnamefont{Hulin}},
  \bibinfo{journal}{Physical review Letters} \textbf{\bibinfo{volume}{90}},
  \bibinfo{pages}{144301} (\bibinfo{year}{2003}).

\bibitem[{\citenamefont{Pusey et~al.}(1989)\citenamefont{Pusey, van Megen,
  Bartlett, Ackerson, Rarity, and Underwood}}]{Pusey1989}
\bibinfo{author}{\bibfnamefont{P.~N.} \bibnamefont{Pusey}},
  \bibinfo{author}{\bibfnamefont{W.}~\bibnamefont{van Megen}},
  \bibinfo{author}{\bibfnamefont{P.}~\bibnamefont{Bartlett}},
  \bibinfo{author}{\bibfnamefont{B.~J.} \bibnamefont{Ackerson}},
  \bibinfo{author}{\bibfnamefont{J.~G.} \bibnamefont{Rarity}},
  \bibnamefont{and} \bibinfo{author}{\bibfnamefont{S.~M.}
  \bibnamefont{Underwood}}, \bibinfo{journal}{Physical Review Letters}
  \textbf{\bibinfo{volume}{63}}, \bibinfo{pages}{2753} (\bibinfo{year}{1989}).

\bibitem[{\citenamefont{Denkov et~al.}(1996)\citenamefont{Denkov, Yoshimura,
  Nagayama, and Kouyama}}]{Denkov1996}
\bibinfo{author}{\bibfnamefont{N.~D.} \bibnamefont{Denkov}},
  \bibinfo{author}{\bibfnamefont{H.}~\bibnamefont{Yoshimura}},
  \bibinfo{author}{\bibfnamefont{K.}~\bibnamefont{Nagayama}}, \bibnamefont{and}
  \bibinfo{author}{\bibfnamefont{T.}~\bibnamefont{Kouyama}},
  \bibinfo{journal}{Physical Review Letters} \textbf{\bibinfo{volume}{76}},
  \bibinfo{pages}{2354} (\bibinfo{year}{1996}).

\bibitem[{\citenamefont{Mueller et~al.}(2010)\citenamefont{Mueller, Llewellin,
  and Mader}}]{Mueller2010}
\bibinfo{author}{\bibfnamefont{S.}~\bibnamefont{Mueller}},
  \bibinfo{author}{\bibfnamefont{E.~W.} \bibnamefont{Llewellin}},
  \bibnamefont{and} \bibinfo{author}{\bibfnamefont{H.~M.} \bibnamefont{Mader}},
  \bibinfo{journal}{Proc. R. Soc. A} \textbf{\bibinfo{volume}{466}},
  \bibinfo{pages}{1201} (\bibinfo{year}{2010}).

\bibitem[{\citenamefont{Ackerson and Pusey}(1988)}]{Ackerson1988}
\bibinfo{author}{\bibfnamefont{B.~J.} \bibnamefont{Ackerson}} \bibnamefont{and}
  \bibinfo{author}{\bibfnamefont{P.~N.} \bibnamefont{Pusey}},
  \bibinfo{journal}{Physical Review Letters} \textbf{\bibinfo{volume}{61}},
  \bibinfo{pages}{1033} (\bibinfo{year}{1988}).

\bibitem[{\citenamefont{Wu et~al.}(2009)\citenamefont{Wu, Derks, van Blaaderen,
  and Imhof}}]{Wu2009}
\bibinfo{author}{\bibfnamefont{Y.~L.} \bibnamefont{Wu}},
  \bibinfo{author}{\bibfnamefont{D.}~\bibnamefont{Derks}},
  \bibinfo{author}{\bibfnamefont{A.}~\bibnamefont{van Blaaderen}},
  \bibnamefont{and} \bibinfo{author}{\bibfnamefont{A.}~\bibnamefont{Imhof}},
  \bibinfo{journal}{PNAS} \textbf{\bibinfo{volume}{106}},
  \bibinfo{pages}{10564} (\bibinfo{year}{2009}).

\bibitem[{\citenamefont{Jerkins et~al.}(2008)\citenamefont{Jerkins, Schr\"oter,
  Swinney, Senden, Saadatfar, and Aste}}]{Jerkins2008}
\bibinfo{author}{\bibfnamefont{M.}~\bibnamefont{Jerkins}},
  \bibinfo{author}{\bibfnamefont{M.}~\bibnamefont{Schr\"oter}},
  \bibinfo{author}{\bibfnamefont{H.~L.} \bibnamefont{Swinney}},
  \bibinfo{author}{\bibfnamefont{T.~J.} \bibnamefont{Senden}},
  \bibinfo{author}{\bibfnamefont{M.}~\bibnamefont{Saadatfar}},
  \bibnamefont{and} \bibinfo{author}{\bibfnamefont{T.}~\bibnamefont{Aste}},
  \bibinfo{journal}{Physical review Letters} \textbf{\bibinfo{volume}{101}},
  \bibinfo{pages}{018301} (\bibinfo{year}{2008}).

\bibitem[{\citenamefont{Francois et~al.}(2013)\citenamefont{Francois,
  Saadatfar, Cruikshank, and Sheppard}}]{Francois2013}
\bibinfo{author}{\bibfnamefont{N.}~\bibnamefont{Francois}},
  \bibinfo{author}{\bibfnamefont{M.}~\bibnamefont{Saadatfar}},
  \bibinfo{author}{\bibfnamefont{R.}~\bibnamefont{Cruikshank}},
  \bibnamefont{and} \bibinfo{author}{\bibfnamefont{A.}~\bibnamefont{Sheppard}},
  \bibinfo{journal}{Physical Review Letters} \textbf{\bibinfo{volume}{111}},
  \bibinfo{pages}{148001} (\bibinfo{year}{2013}).

\bibitem[{\citenamefont{Bernal}(1964)}]{Bernal1964}
\bibinfo{author}{\bibfnamefont{J.~D.} \bibnamefont{Bernal}},
  \bibinfo{journal}{Proc. R. Soc. A} \textbf{\bibinfo{volume}{280}},
  \bibinfo{pages}{299} (\bibinfo{year}{1964}).

\bibitem[{\citenamefont{Anikeenko and Medvedev}(2007)}]{Anikeenko2007}
\bibinfo{author}{\bibfnamefont{A.~V.} \bibnamefont{Anikeenko}}
  \bibnamefont{and} \bibinfo{author}{\bibfnamefont{N.~N.}
  \bibnamefont{Medvedev}}, \bibinfo{journal}{Physical Review Letters}
  \textbf{\bibinfo{volume}{98}}, \bibinfo{pages}{235504}
  (\bibinfo{year}{2007}).

\bibitem[{\citenamefont{Anikeenko et~al.}(2008)\citenamefont{Anikeenko,
  Medvedev, and Aste}}]{Anikeenko2008}
\bibinfo{author}{\bibfnamefont{A.~V.} \bibnamefont{Anikeenko}},
  \bibinfo{author}{\bibfnamefont{N.~N.} \bibnamefont{Medvedev}},
  \bibnamefont{and} \bibinfo{author}{\bibfnamefont{T.}~\bibnamefont{Aste}},
  \bibinfo{journal}{Physical Review E} \textbf{\bibinfo{volume}{77}},
  \bibinfo{pages}{031101} (\bibinfo{year}{2008}).

\bibitem[{\citenamefont{Sch\"ope et~al.}(2006)\citenamefont{Sch\"ope, Bryant,
  and van Megen}}]{Schope2006}
\bibinfo{author}{\bibfnamefont{H.~J.} \bibnamefont{Sch\"ope}},
  \bibinfo{author}{\bibfnamefont{G.}~\bibnamefont{Bryant}}, \bibnamefont{and}
  \bibinfo{author}{\bibfnamefont{W.}~\bibnamefont{van Megen}},
  \bibinfo{journal}{Physical Review E} \textbf{\bibinfo{volume}{74}},
  \bibinfo{pages}{060401} (\bibinfo{year}{2006}).

\bibitem[{\citenamefont{Martin et~al.}(2003)\citenamefont{Martin, Bryant, and
  van Megen}}]{Martin2003}
\bibinfo{author}{\bibfnamefont{S.}~\bibnamefont{Martin}},
  \bibinfo{author}{\bibfnamefont{G.}~\bibnamefont{Bryant}}, \bibnamefont{and}
  \bibinfo{author}{\bibfnamefont{W.}~\bibnamefont{van Megen}},
  \bibinfo{journal}{Physical Review E} \textbf{\bibinfo{volume}{67}},
  \bibinfo{pages}{061405} (\bibinfo{year}{2003}).

\end{thebibliography}

\end{document}